\documentclass[12pt]{iopart}
\usepackage{iopams}  
\usepackage{setstack}

\begin{document}
\title[Asymptotic diagonalization of the Discrete Lax pair]{Asymptotic diagonalization of the Discrete Lax pair around singularities and conservation laws for dynamical systems}

\author{I T Habibullin$^{1,2}$ and M N Poptsova$^1$}

\address{$^1$Ufa Institute of Mathematics, Russian Academy of Sciences,
112 Chernyshevsky Street, Ufa 450008, Russian Federation}
\address{$^2$Bashkir State University, 32 Validy Street, Ufa 450076 , Russian Federation}
\eads{\mailto{habibullinismagil@gmail.com} and \mailto{mnpoptsova@gmail.com}}

\begin{abstract}
A method of the formal diagonalization of the discrete linear operator with a parameter is studied. In the case when the operator provides a Lax operator for a nonlinear quad system the formal diagonalization method allows one to describe effectively conservation laws and generalized symmetries for this system. Asymptotic representation of the Lax operators eigenfunctions are constructed and infinite series of conservation laws are described for the quad system connected with $A^{(1)}_3$ affine Lie algebra, for the modified discrete Boussinesq system and for the discrete Tzitzeica equation. For a newly found multiquadratic discrete model conservation laws and several generalized  symmetries are presented.

\end{abstract}
\submitto{J. Phys. A: Math. Theor. 48 (2015) 115203.}
\maketitle

\section{Introduction}

In \cite{HabLOMI,HabYang} a method of asymptotic diagonalization of the Lax pairs associated with nonlinear discrete and semidiscrete models was suggested allowing to describe conservation laws and higher symmetries for the corresponding nonlinear models. Efficiency of the method was approved by application to numerous examples in \cite{HabYang,GY14,GHY14}. 

In this article we adopt the algorithm to a large class of the linear systems having more complicated analytical structure than those studied earlier.

\eqnobysec
We consider a discrete linear equation of the form
\begin{equation}  \label{eq1}
y_{n+1}(\lambda) = f_n(u,\lambda) y_n( \lambda),
\end{equation}
where the potential $f = f_n(u, \lambda) \in \mathbb{C}^{N\times N}$ depends on the integer $n \in (-\infty, +\infty)$, on a functional parameter $u = u_n$ and its shifts  called the dynamical variables and the complex parameter $\lambda$. The potential is a meromorphic function of $\lambda$, where $\lambda$ ranges a domain $U\subset {\mathbb C}$. Important properties of the equation are connected with the singularities of $f_n(u, \lambda)$.  We call $\lambda = \lambda_0$ a singular point if at least one of the functions $f_n(u, \lambda)$, $f^{-1}_n(u,\lambda)$ has a pole at this point. It is supposed that $\lambda_0$ does not depend on $n$. Emphasize that some of the singularities can be removed by an appropriately chosen gauge transformation, where the matrix-valued function $r_n(\lambda) \in \mathbb{C}^{N\times N}$ does not depend on the functional parameter $u$ (in some cases it is easily found)
\begin{equation}	\label{eq2}
y_n = r_n(\lambda) \tilde{y}_n
\end{equation}
combined with the change of the parameter $\lambda=\omega(\tilde \lambda)$.
Transformation (\ref{eq2}) reduces equation (\ref{eq1}) to the form
\begin{equation}	\label{eq3}
\tilde{y}_{n+1} = \tilde{f}_n(u,\lambda) \tilde{y}_n
\end{equation}
with the new potential $\tilde{f}_n(u,\lambda) = r^{-1}_{n+1}(\lambda) f_n(u,\lambda)r_n(\lambda)$.

As an illustrative example consider the well-known linear equation associated in the context of the integrability with $\mathrm{H1}$ equation from the ABS list \cite{ABS}. It is an equation of the form (\ref{eq1}) with the potential \cite{nc95}
\begin{equation}  	\label{eq4}
f_n(u,\lambda) = \left( \begin{array}{cc}
-u_{n+1} & 1\\
-\lambda -u_n u_{n+1} & u_n
\end{array} \right),
\end{equation}
where $u=u_n$ is an arbitrary function. Evidently equation (\ref{eq1}), (\ref{eq4}) has two singular points: $\lambda = 0$ and $\lambda = \infty$ since $\det f=\lambda$. However $\lambda = \infty$ is removed by the transformation (\ref{eq3}), where
\begin{equation}	\label{eq5}
r_n(\lambda) = \left( \begin{array}{cc}
\xi^n & 0\\
0 & \xi^{n+1}
\end{array} \right), \quad \xi=\sqrt{\lambda}.
\end{equation}
Indeed for this case the new potential
\begin{equation*}	
\tilde{f}_n(\xi) = \left( \begin{array}{cc}
-u_{n+1} \xi^{-1} & 1\\
-1 - u_nu_{n+1} \xi^{-2} & u_n\xi^{-1}
\end{array} \right)
\end{equation*}
has the only singularity point $\xi = 0$.

The key step of the diagonalization algorithm is to reduce system (\ref{eq1}) to some special form around the singular point $\lambda=\lambda_0$
\begin{equation}\label{SpecialForm}
\psi_{n+1}(\lambda)=P_n(\lambda)Z\psi_n(\lambda)
\end{equation}
where $P_n(\lambda)$ together with $P^{-1}_n(\lambda)$ are analytic around $\lambda_0$, the leading principal minors of $P_n(\lambda)$ satisfy additional conditions (\ref{eq14}) and $Z$ is of the form (\ref{eq13}).
Remark that we failed to prove that the special form (\ref{SpecialForm}) exists around arbitrary unremovable singular point of the equation (\ref{eq1}). However we have a general scheme which provides as a rule transition from   equation (\ref{eq1}) to (\ref{SpecialForm}).

{\bf Example.} Explain the scheme briefly with the example (\ref{eq1}), (\ref{eq4}) around the unremovable singular point $\lambda=0$. Due to the Gauss method we have a decomposition of the potential $f$ as a product of triangular matrices
\begin{equation}\label{gauss}
f=f_Lf_U
\end{equation}
where the lower-triangular and upper-triangular factors are defined as follows
\begin{equation}	\label{factors}
f_L = \left( \begin{array}{cc} 1 & 0\\
\displaystyle u_n+\frac{\lambda}{u_{n+1}} & \displaystyle \frac{-1}{u_{n+1}}\end{array} \right), \qquad 
f_U = \left( 
\begin{array}{cc}
 -u_{n+1} & 1\\
0 &  \lambda 
\end{array} \right).
\end{equation}
The first factor in (\ref{gauss}) is bounded and non-degenerate around $\lambda=0$. However the second factor is degenerate. But  $\beta_n=Z^{-1}f_U$ is bounded and non-degenerate for $\lambda=0$ if the diagonal factor is chosen as $Z=\mathrm{diag}(1,\lambda)$. Thus $f$ is represented as $f=f_LZ\beta_n$ and the singular part of $f$ at $\lambda=0$ is completely localized in $Z$.

Now change the variables in (\ref{eq1}), (\ref{eq4}) $\psi_n=\beta_ny_n$ and find the dezired form (\ref{SpecialForm}) with $P_n=\beta_{n+1}f_L$.
Let us give $P_n$ explicitly
\begin{equation}  	\label{eq44}
P_n(u,\lambda) = \left( \begin{array}{cc}
\displaystyle u_n-u_{n+2}+\frac{\lambda}{u_{n+1}} & \displaystyle -\frac{1}{u_{n+1}}\\
\displaystyle u_n+\frac{\lambda}{u_{n+1}} & \displaystyle -\frac{1}{u_{n+1}}
\end{array} \right).
\end{equation}

It follows from the explicit formula that both the determinant 
$\displaystyle \det P_n(\lambda=0)=\frac{u_{n+1}}{u_n}$ and  $\det_1P_n(\lambda=0)=u_{n}- u_{n-2}$
do not vanish if the conditions hold 
\begin{equation}\label{inequalities}
\forall n \,u_n\neq 0, \quad u_n\neq u_{n+2}.
\end{equation}
Here and everywhere below $\det_j A$ denotes the $j$-th order leading principal minor of the matrix $A$ (i.e., it consists of matrix elements in rows and columns from $1$ to $j$). Thus all the settings of the Theorem 1 below on the formal diagonalization are satisfied. Note that the inequalities (\ref{inequalities}) are dictated not only by the  formal diagonalization technique they have a deeper meaning. The conserved densities and symmetries for the equation $\mathrm{H1}$ are correctly defined only under these conditions and similar conditions on the other direction.

In some cases the problem of reducing system (\ref{eq1}) to the form (\ref{SpecialForm}) might cause difficulties. That puts a potential  limitation to apply the method.
  
There are several methods for constructing conservation laws for the quad equations: the direct method suggested in \cite{rh07_1}, the Gardner method \cite{Ras}, the method of the canonical densities, connected with  the symmetries and the recursion operators \cite{mwx,xn12,ly09,ly11,ggh11} and so on. A method based on reducing the Lax equation to the Riccati type nonlinear system is suggested in \cite{Cheng1,Cheng2}.

Formal diagonalization as an implement for constructing conservation laws is more preferable when applied to the potentials $f$, $g$ either having large matrix dimension or being high degree polynomials on $\lambda$. The algorithm is very simple to apply since it does not need any additional construction only the Lax pair reduced to the special form. Some of the advantages of the
formal diagonalization method have successfully been demonstrated in the recent article \cite{GHY14} where the conservation laws and symmetries were constructed to nonautonomous  equations.

As it has been observed in \cite{Dri} the formal diagonalization of the Lax pair is an algebraic background of the inverse scattering transform method. Actually formula (\ref{eq10}) below defines a formal asymptotic expansion of the solution to the direct scattering problem. Note that the ``phase'' and the ``amplitude'' in (\ref{eq10}) are connected respectively with the integrals of motion and the symmetries of the corresponding nonlinear quad equation. An open problem is to apply the expansion (\ref{eq10}) in order to construct explicit particular solutions of the quad equation via quasi-classical approximation. Thus the research is undoubtedly in demand.

The behavior asymptotic in a parameter of linear differential equations has been well studied. A similar problem for the discrete linear systems has got less attention. One of the purposes of the present article is to continue the investigation of the discrete linear systems started in \cite{HabLOMI}, \cite{HabYang}, where the particular case is considered when the  matrix $Z$ in the special form (\ref{SpecialForm}) has only simple eigenvalues, more precisely, in the formula (\ref{eq13}), $r=N$. It turned out that there are discrete models (see, (\ref{discreteKP}) for $N' > 1$, (\ref{Tz:eq1}) and (\ref{S6eq2dashed}) below) for which the Lax pairs are reduced to the form (\ref{SpecialForm}) having the factor $Z$ with multiple eigenvalues, i.e. $r<N$.

Briefly explain how the article is organized. In \S2 the theorem on asymptotic diagonalization is proved.  An algorithm of constructing conservation laws via diagonalization is discussed in \S3. In \S4-5 the diagonalization method is applied to some specific models like the discrete mKdV, the discrete modified Boussinesq system, the discrete system, connected with the affine Lie algebra $A^{(1)}_3$, the Tzitzeica equation found in \cite{Adler,Bobenko-Schief}.

In \S6 by the Zakharov-Shabat dressing method \cite{ZakharovShabat} a discrete model is found 
\begin{eqnarray}
\fl 2 \varepsilon^2 u u_{0,1}(
u_{1,0}  u_{1,1} +  u u_{1,0} + u_{0,1} u_{1,1} + u u_{0,1})-4 \varepsilon^2 u u_{0,1} (u_{1,0} u_{0,1} + u u_{1,1})-\nonumber\\ 
 -  \varepsilon^4  u^2_{0,1}(u_{1,1}-u)^2 + 2 u^2 u_{1,0} u_{0,1} - u^2 (u^2_{1,0} + u^2_{0,1}) = 0	\label{S1eq2}
\end{eqnarray}
which possibly is new.  Here $\varepsilon$  is a constant parameter and as usually the subscripts mean the shifts of the arguments: $u_{k,s}=u(n+k,m+s)$. But sometimes subscripts denote also an element of the matrix. This should not lead to confusion, because it is always clear from the context what is meant. Equation (\ref{S1eq2}) is multi-quadratic since the left hand side is a quadratic polynomial with respect to any of the variables $u$, $u_{1,0}$, $u_{0,1}$, $u_{1,1}$. Various examples of multi-quadratic equations appeared in \cite{HietarintaBou}-\cite{Lewin}. A large class of such equations is systematically investigated in \cite{AtkinNies}. Emphasize that equation (\ref{S1eq2}) does not belong to the class studied earlier. It can be rewritten in a short form
$$\fl \varepsilon\sqrt{(u_{1,1}-u_{0,1})u_{0,1}} = \sqrt{(u_{1,0}-u)u} 
-\sqrt{(1+\varepsilon^{2})u u_{0,1} -  u_{\vphantom{1}}^2 - \varepsilon^{2}u^2_{0,1}}.$$
It is shown that the model admits an infinite series of conservation laws, several of them are given in explicit form. Generalized symmetries of low orders  are constructed.

Some explanation how to reduce a discrete linear operator to the special form is given in the Appendix.

\section{Asymptotic diagonalization of the discrete operator around a singular point}

Suppose that the potential $f_n(u,\lambda)$ has a pole at the point $\lambda = \lambda_0$. Then we have the following Laurent expansion around this point
\begin{equation}	\label{eq6}
f_n(u,\lambda) = (\lambda - \lambda_0)^{-k} f^{(k)}(n) + (\lambda - \lambda_0)^{-k+1} f^{(k-1)}(n) + \cdots, \quad k\geq1.
\end{equation}
In this section we explain the algorithm of the formal diagonalization of the equation (\ref{eq1}) at a vicinity of the point $\lambda = \lambda_0$. Recall that  equation (\ref{eq1}) is diagonalized if there exist two formal series 
\begin{equation}	\label{eq7}
R(n, \lambda) = R^{(0)} + R^{(1)} (\lambda - \lambda_0) + R^{(2)} (\lambda - \lambda_0)^2 +\cdots, 
\end{equation}
and 
\begin{equation}	\label{eq8}
h(n,\lambda) = h^{(0)} + h^{(1)}(\lambda - \lambda_0) + h^{(2)} (\lambda - \lambda_0)^2 + \cdots
\end{equation}
with the matrix coefficients $R^{(j)}, h^{(j)} \in \mathbb{C}^{N\times N}$, where $h^{(j)}$ for $\forall j$ is a diagonal (block-diagonal) matrix such that the formal change of the dependent variables $y = R \varphi$ reduces equation (\ref{eq1}) to the form
\begin{equation} 	\label{eq9}
\varphi_{n+1} = h Z \varphi_n,
\end{equation}
here $Z = (\lambda - \lambda_0)^d$, and $d \in \mathbb{C}^{N\times N}$ is a diagonal matrix with entire entries. It is supposed that $\det R^{(0)} \neq 0$, $\det h^{(0)} \neq 0$. It follows from (\ref{eq7})--(\ref{eq9}) that the diagonalizable equation (\ref{eq1}) admits a formal solution given by the following asymptotic expansion
\begin{equation} \label{eq10}
y_n(\lambda) = R(n, \lambda) e^{\sum_{s = n_0}^{n-1} \log h(s, \lambda)} Z^n
\end{equation}
with the ``amplitude'' $A=R(n, \lambda)$ and ``phase'' $\phi=n\log Z+\sum_{s = n_0}^{n-1} \log h(s, \lambda)$.

\textbf{Proposition.} If the condition $\det f^{(k)} (n) \neq 0$  holds for $\forall n$ in the decomposition (\ref{eq6}) then the singularity at the point $\lambda = \lambda_0$ is removable.

Proof. Let us apply transformation (\ref{eq2}) with $r_n(\lambda) = (\lambda - \lambda_0)^{-nk}$ to (\ref{eq1}) then we get $\tilde{f}_n(\lambda_0) = (\lambda - \lambda_0)^k f_n(\lambda)$, which is bounded for $\lambda \rightarrow \lambda_0$ and\\$\det \tilde{f}_n(\lambda_0) = \det f^{(k)}(n) \neq 0$. Thus $\lambda_0$ is not a singular point for $\tilde{f}_n(\lambda)$. \fullsquare

Suppose that the potential $f_n(u,\lambda)$ is represented as follows 
\begin{equation} 	\label{eq11}
f_n(u, \lambda) = \alpha_n(u,\lambda) Z \beta_n(u,\lambda),
\end{equation}
where $\alpha_n(u, \lambda), \beta_n(u, \lambda) \in \mathbb{C}^{N\times N}$ are analytic and non-degenerate around $\lambda=\lambda_0$, $Z$ is a diagonal matrix of the form
\begin{equation}		\label{eq13}
Z = \left( \begin{array}{cccc}
(\lambda-\lambda_0)^{\gamma_1} E_1 & 0 & \ldots & 0\\
0 & (\lambda-\lambda_0)^{\gamma_2} E_2 & \ldots & 0\\
\vdots & \vdots & \ddots & \vdots \\
0 & 0 & \ldots & (\lambda-\lambda_0)^{\gamma_r} E_r
\end{array} \right),
\end{equation} 
where $\forall j$ $E_j$ is the unity matrix of the size $e_j\times e_j$, $e_j<N$ and the exponents are pairwise different integers $\gamma_1 < \gamma_2 < \ldots < \gamma_r$. Evidently $\sum e_i=N$. 
Usually the problem of finding the factors $\alpha$, $\beta$, and $Z$ is solved by using the Gauss factorization of the potential $f$ by triangular matrices (see \cite{HabYang}). The main idea is illustrated with Example  in Introduction. More complicated examples are considered in \S4-\S6. Whenever the decomposition (\ref{eq11}) is found one makes in the equation $y_{n+1}=\alpha_n(u,\lambda) Z \beta_n(u,\lambda)y_n$ the change of the variables $\psi_n=\beta_n (u,\lambda)y_n$ reducing equation (\ref{eq1}) to the following form
\begin{equation}\label{MainSpecialForm}
\psi_{n+1}=P_n(u, \lambda)Z\psi_n,\quad \mbox{where}\quad P_n(u, \lambda) = D_n\left(\beta_n(u, \lambda)\right) \alpha_n(u,\lambda)
\end{equation}
and the shift operator acts due to the rule $D_ng_{n}=g_{n+1}$.

Let us formulate our main requirement to the factorization (\ref{eq11}):  suppose that the leading principal minors satisfy the conditions
\begin{equation} \label{eq14}
\det_j P_n(u, \lambda_0) \neq 0  \mbox{ for } j= e_1, e_1+e_2, e_1+e_2+e_3, \ldots, N.
\end{equation}

Obviously determinant of the function (\ref{eq6}) is of the form 
\begin{equation*}
\det f_n(u,\lambda)=(\lambda-\lambda_0)^{k_o}(c_0+c_1(\lambda-\lambda_0)+\dots),
\end{equation*}
where $k_0$ is an integer and $c_0\neq0$. By evaluating the determinant of both sides in the equation $P_n(u,\lambda)Z=D_n(\beta_n(u,\lambda))f_n\beta_n(u,\lambda)$ and comparing the singularities one easily obtains  $(\lambda-\lambda_0)^{\gamma_1+\gamma_2+...\gamma_r}=(\lambda-\lambda_0)^{k_0}$ since the functions $P_n(u,\lambda))$ and $\beta_n(u,\lambda)$ are regular and nondegenerate at $\lambda_0$. Therefore we have $\sum \gamma_i=k_0$.

Turn back to the problem of diagonalization. Concentrate first on the equation (\ref{MainSpecialForm}) as due to the relation $\psi_n=\beta_n (u,\lambda)y_n$ diagonalization of this equation implies immediately diagonalization of (\ref{eq1}). 

\textbf{Theorem 1.} Suppose that equation (\ref{MainSpecialForm}) satisfies (\ref{eq13}), (\ref{eq14}). Then there exist formal series 
\begin{equation} \label{S1eq14}
\eqalign{T(n,\lambda) = \sum_{i \geq 0}^{\infty} T^{(i)} (\lambda - \lambda_0)^i, \qquad  h(n, \lambda) = \sum_{i \geq 0}^{\infty} h^{(i)} (\lambda - \lambda_0)^i}
\end{equation}
where $\det h^{(0)} \neq 0$, $\det T^{(0)} \neq 0$ and $h$ has the following block-diagonal structure
\begin{equation} \label{eq14dashed}
h = \left( \begin{array}{cccc}
h_{11} & 0 & \dots & 0\\
0 & h_{22} & \dots & 0\\
\vdots & \vdots & \ddots & \vdots\\
0 & 0 & \dots & h_{rr} 
\end{array} \right),
\end{equation}
where $h_{jj}$ is a quadratic matrix of the size $e_j \times e_j$ such that the formal change of the variables $\psi=T\varphi$ reduces (\ref{MainSpecialForm}) to the diagonal form (\ref{eq9}). The series $T$ is defined up to multiplication by a formal series with block-diagonal members having the same block structure as $h$ and we can always choose them in such a way that all the coefficients $T^{(i)}, h^{(i)} \in \mathbb{C}^{N\times N}$ are dependent on a finite set (depending on $i$) of the dynamical variables ${u(k)}_{k=\overline{-\infty,\infty}}$.

Proof. By substituting $\psi=T\varphi$ into (\ref{MainSpecialForm}) one gets $D_n(T)D_n(\varphi)=PZ\varphi$. It follows from (\ref{eq9}) that $D_n(\varphi) \varphi^{-1}=hZ$. Therefore the former gives an equation for defining the unknown $T$ and $h$:
\begin{equation}	\label{eq16}
D_n(T) h = P_n(u, \lambda) \overline{T}, \quad\mbox{with} \quad \overline{T} = Z T Z^{-1}.
\end{equation}
Since $P_n(u,\lambda)$ is analytic around $\lambda_0$ we have  $P_n(u, \lambda) = \sum_{i \geq 0}^{\infty} P^{(i)} (\lambda - \lambda_0)^i$. By construction $\overline T$ is a formal power series. It follows from the equation\\$\overline T= P^{-1}_n(u, \lambda) D_n(T) h$ and the inequality $\det P_n(u,\lambda_0)\neq0$ that the series $\overline T$ does not contain terms with negative powers of $\lambda-\lambda_0$
\begin{equation}	\label{eq18}
\overline{T}  = \sum_{i \geq 0}^{\infty} \overline{T}^{(i)} (\lambda - \lambda_0)^i.
\end{equation} 

In what follows we will use the block representation of $N\times N$ matrices. The block element $A_{ij}$ of the matrix $A$ has $e_i$ rows and $e_j$ columns. It is evident that $Z h = h Z$ due to (\ref{eq13}) and (\ref{eq14dashed}). The conjugation transformation $T \rightarrow Z T Z^{-1}$ can be expressed in an explicit block form
\begin{equation}	\label{eq19}
\fl Z T Z^{-1} 
= \left( \begin{array}{cccc}
 T_{11} & (\lambda-\lambda_0)^{\gamma_{12}} T_{12} & \ldots &(\lambda-\lambda_0)^{\gamma_{1r}} T_{1r}\\
(\lambda-\lambda_0)^{\gamma_{21}} T_{21} & \ T_{22} & \ldots & (\lambda-\lambda_0)^{\gamma_{2r}} T_{2r}\\
\vdots & \vdots & \ddots & \vdots\\
(\lambda-\lambda_0)^{\gamma_{r1}} T_{r1} & (\lambda-\lambda_0)^{\gamma_{r2}} T_{r2} & \ldots &  T_{rr}
\end{array} \right),
\end{equation}
where $\gamma_{ij} = \gamma_i - \gamma_j$. Formula (\ref{eq19}) shows that the block entry $\overline{T}_{ij}$ of the matrix $\overline{T}$ is given by $\overline{T}_{ij} = (\lambda-\lambda_0)^{\gamma_{ij}} T_{ij}$ where $T_{ij}$ is the block element of $T$. For these block elements we have formal series (see (\ref{S1eq14}), (\ref{eq18}) above)
\begin{eqnarray*}
T_{ij}(\lambda) = T^{(0)}_{ij} + (\lambda - \lambda_0) T^{(1)}_{ij} + \cdots, \\ \overline{T}_{ij}(\lambda) = \overline{T}^{(0)}_{ij} + (\lambda - \lambda_0) \overline{T}^{(1)}_{ij} + \cdots
\end{eqnarray*}
which satisfy due to (\ref{eq19}) the following equation
\begin{equation}	\label{eq20}
\overline{T}^{(0)}_{ij} + \overline{T}^{(1)}_{ij}(\lambda - \lambda_0) + \cdots = T^{(0)}_{ij}(\lambda-\lambda_0)^{\gamma_{ij}} + T^{(1)}_{ij} (\lambda - \lambda_0)^{\gamma_{ij} + 1} + \cdots
\end{equation}
By comparing the coefficients in (\ref{eq20}) one finds
\begin{eqnarray}
{T\vphantom{\overline{T}}}^{(p)}_{ii} = \overline{T}^{(p)}_{ii}, 	\label{eq21}\\
{T\vphantom{\overline{T}}}^{(p)}_{ij} = \overline{T}^{(p + \gamma_i - \gamma_j)}_{i j} \qquad \mbox{for $i<j$},	\label{eq22}\\
{T\vphantom{\overline{T}}}^{(p)}_{ij} = \overline{T}^{(p + \gamma_j - \gamma_i)}_{i j} \qquad \mbox{for $i>j$}.	\label{eq23}
\end{eqnarray}
It is supposed in (\ref{eq21})--(\ref{eq23}) that ${T\vphantom{\overline{T}}}^{(p)}_{ij} = \overline{T}^{(p)}_{ij} = 0$ for $p<0$. It follows from (\ref{eq21})--(\ref{eq23}) that $T^{(0)}$ is a lower block triangular matrix. Indeed, we have for $i < j$ that $T^{(0)}_{ij} = T^{(\gamma_i - \gamma_j)}_{ij}$ is zero, because $\gamma_i - \gamma_j < 0$. Similarly $\overline{T}^{(0)}_{ij} = 0$ for $i>j$ therefore $\overline{T}^{(0)}$ is upper block triangular.

Let us substitute series (\ref{S1eq14}), (\ref{eq18}) into the equation (\ref{eq16})
\begin{eqnarray*}
\fl D_n \left( \sum_{i \geq 0} {T\vphantom{\overline{T}}}^{(i)} (\lambda - \lambda_0)^i \right) \sum_{i \geq 0} {h\vphantom{\overline{T}}}^{(i)} (\lambda - \lambda_0)^i \\= \sum_{i \geq 0} {P\vphantom{\overline{T}}}^{(i)} (\lambda - \lambda_0)^i \sum_{i \geq 0} \overline{T}^{(i)} (\lambda - \lambda_0)^i
\end{eqnarray*}
and collect the coefficients before the powers of $\lambda - \lambda_0$. It gives a sequence of equations:
\begin{eqnarray}
D_n\left({T\vphantom{\overline{T}}}^{(0)}\right) {h\vphantom{\overline{T}}}^{(0)} = {P\vphantom{\overline{T}}}^{(0)} \overline{T}^{(0)},	\label{eq24}\\
D_n\left({T\vphantom{\overline{T}}}^{(0)}\right) {h\vphantom{\overline{T}}}^{(1)} + D_n\left({T\vphantom{\overline{T}}}^{(1)}\right) {h\vphantom{\overline{T}}}^{(0)} = {P\vphantom{\overline{T}}}^{(0)} \overline{T}^{(1)} + {P\vphantom{\overline{T}}}^{(1)} \overline{T}^{(0)},	\label{eq25}\\
\sum_{k=0}^s D_n\left({T\vphantom{\overline{T}}}^{(k)}\right) {h\vphantom{\overline{T}}}^{(s-k)} = \sum_{k=0}^s {P\vphantom{\overline{T}}}^{(k)} \overline{T}^{(s-k)}, \qquad s \geq 2. \label{eq26}
\end{eqnarray} 
Concentrate on the first equation. This equation is easily reduced to the Gauss problem of decomposition of the matrix $P^{(0)}$ into a product $D_n\left({T\vphantom{\overline{T}}}^{(0)}\right) {h\vphantom{\overline{T}}}^{(0)} \left(\overline{T}^{(0)}\right)^{-1}$ of three matrices -- the lower block-triangular $D_n\left({T\vphantom{\overline{T}}}^{(0)}\right)$, the block diagonal ${h\vphantom{\overline{T}}}^{(0)}$ and the upper block-triangular $\overline{T}^{(0)}$. The regularity condition (\ref{eq14}) guarantees its solvability. In addition we assume that diagonal blocks $T_{ii}$ and $\overline{T}_{ii}$ are equal to the unity matrix $E_i$ of the size $e_i \times e_i$. Evidently this provides the unique solvability of equation (\ref{eq24}). 

There is a freedom in choosing the coefficients of the series $T$. For the sake of simplicity we assume that the diagonal blocks of the matrices ${T\vphantom{\overline{T}}}^{(k)}$ and $\overline{T}^{(k)}$ vanish for all $k \geq 1$ and expand each of the matrices ${T\vphantom{\overline{T}}}^{(k)}$ and $\overline{T}^{(k)}$ for $k \geq 1$ into the sum of block lower- and block upper-triangular matrices with zero diagonal blocks:
\begin{equation}	\label{eq27}
{T}_{\vphantom{L}}^{(k)} = T^{(k)}_{L} + T^{(k)}_{U}, \qquad \overline{T}^{(k)} = \overline{T}^{(k)}_{L} + \overline{T}^{(k)}_{U}.
\end{equation}
One can easily find the matrices ${T\vphantom{\overline{T}}}^{(1)}_{U}$ and $\overline{T}^{(1)}_{ L}$. Indeed, if $\gamma_i - \gamma_j = -1$ for $i < j$ then we obtain ${T\vphantom{\overline{T}}}^{(1)}_{ij} = \overline{T}^{(0)}_{ij}$ from (\ref{eq22}), we found this element in the previous step. If $\gamma_i - \gamma_j \leq -2$ then we obtain ${T\vphantom{\overline{T}}}^{(1)}_{ij} = \overline{T}^{(1+\gamma_i - \gamma_j)}_{ij} = 0$ from (\ref{eq22}) because the first index $1 + \gamma_i - \gamma_j$ is negative. Therefore we already know the matrix $T^{(1)}_{U}$. In a similar way we find also $\overline{T}^{(1)}_{L}$.

To find the unknown $T^{(1)}_{L}$ and $\overline{T}^{(1)}_{U}$ we use the equation (\ref{eq25}) rewritten in the form
\begin{equation*}
{h\vphantom{\overline{T}}}^{(1)} \left({h\vphantom{\overline{T}}}^{(0)}\right)^{-1} + D_n \left( \left({T\vphantom{\overline{T}}}^{(0)}\right)^{-1} {T\vphantom{\overline{T}}}^{(1)}_{L} \right) - {h\vphantom{\overline{T}}}^{(0)} \left(\overline{T}^{(0)}\right)^{-1} \overline{T}^{(1)}_{ U} \left({h\vphantom{\overline{T}}}^{(0)}\right)^{-1} = H_1
\end{equation*}
where the right hand side 
\begin{eqnarray*}
\fl H_1 = D_n\left(\left({T\vphantom{\overline{T}}}^{(0)}\right)^{-1}\right){P\vphantom{\overline{T}}}^{(1)}\overline{T}^{(0)} \left({h\vphantom{\overline{T}}}^{(0)}\right)^{-1} \\
- D_n\left( \left({T\vphantom{\overline{T}}}^{(0)}\right)^{-1} {T\vphantom{\overline{T}}}^{(1)}_{U}\right)
+ {h\vphantom{\overline{T}}}^{(0)} \left(\overline{T}^{(0)}\right)^{-1} \overline{T}^{(1)}_{L}\left({h\vphantom{\overline{T}}}^{(0)}\right)^{-1}
\end{eqnarray*}
contains known matrices. To find the unknowns ${h\vphantom{\overline{T}}}^{(1)}$, ${T\vphantom{\overline{T}}}^{(1)}_{L}$, $\overline{T}^{(1)}_{U}$ we have to expand the matrix $H_1$ into the sum of three summands: block diagonal ${h\vphantom{\overline{T}}}^{(1)} \left({h\vphantom{\overline{T}}}^{(0)}\right)^{-1}$, the block lower-triangular $D_n \left( \left({T\vphantom{\overline{T}}}^{(0)}\right)^{-1} {T\vphantom{\overline{T}}}^{(1)}_{L}\right) $ and the block upper-triangular $-h^{(0)} \left(\overline{T}^{(0)}\right)^{-1} \overline{T}^{(1)}_{ U} \left({h\vphantom{\overline{T}}}^{(0)}\right)^{-1} $. Continuing this way we find all coefficients $T^{(k)}$ and $h^{(k)}$ from the equation 
\begin{equation*}
{h\vphantom{\overline{T}}}^{(k)} \left({h\vphantom{\overline{T}}}^{(0)}\right)^{-1} + D_n \left( \left({T\vphantom{\overline{T}}}^{(0)}\right)^{-1} {T\vphantom{\overline{T}}}^{(k)}_{L}\right) - {h\vphantom{\overline{T}}}^{(0)} \left(\overline{T}^{(0)}\right)^{-1} \overline{T}^{(k)}_U \left({h\vphantom{\overline{T}}}^{(0)}\right)^{-1} = H_k,
\end{equation*}
where the term $H_k$ contains terms found in the preceding steps. 

Suppose that equation (\ref{MainSpecialForm}) can be reduced to one more diagonal form $\tilde\varphi_1 = \tilde h Z \tilde\varphi$ by applying the change of the variables $\psi=\tilde T\tilde\varphi$ then we have $\tilde T\tilde\varphi=T\varphi$ or $\tilde T=TS$ where $S=\varphi^{-1}\tilde\varphi$ is a power series with block-diagonal coefficients. 
\fullsquare

\textbf{Remark.} Note that in order to find the first term of the formal expansion one needs to solve a nonlinear equation (see eq. (\ref{eq24})). It is not the case for the linear differential equations \cite{Wasow}, where all of the coefficients of the corresponding formal series are found by solving linear equations.

\textbf{Corollary.} Linear equation (\ref{eq1}) is reduced to the block-diagonal form (\ref{eq9}) by the following change of the variables $y=R\varphi$ where $R=\beta^{-1}T$.

\section{Asymptotic diagonalization of the Lax operator and conservation laws}

We study a quad system of the form
\begin{equation}	\label{S3eq1}
F (D_m D_n v, D_m v, D_n v, v) = 0,
\end{equation}
where the sought function $v = v(n,m) = \left( v_1(n,m), v_2(n,m), \ldots, v_N(n,m)  \right)$ depends on two integers $n$ and m and the shift operators $D_n$ and $D_m$ act due to the rules $D_n y(n,m) = y(n+1,m)$ and $D_m y(n,m) = y(n,m+1)$. We assume that equation (\ref{S3eq1}) admits a Lax representation, i.e. it is the consistency condition of a pair of linear equations 
\begin{equation}	\label{S3eq2}
\eqalign{
y_{n+1,m} = P_{n,m}(v, \lambda)Z y_{n,m},\\
y_{n,m+1} = R_{n,m}(v,\lambda)y_{n,m}.}
\end{equation}
Here the functions $P_{n,m}(v,\lambda)$, $R_{n,m}(v,\lambda)$ depend on $v$ and a finite numbers of its shifts $D^k_n v$ and $D^k_m v$. The consistency condition of the equations system  (\ref{S3eq2}) can be written as
 \begin{equation}	\label{S3eq3}
 D_m(P) Z R = D_n(R)PZ.
 \end{equation}
Let us introduce the operatots $L = D^{-1}_n PZ$ and $M = D^{-1}_m R$ then equation (\ref{S3eq3}) can be rewritten as
\begin{equation}
[L, M] = 0	\label{S3eq4}
\end{equation}
 
 Note that the first equation (\ref{S3eq2}) is of the form (\ref{MainSpecialForm}). We assume that function $P_{n,m}(v,\lambda)$ satisfies all the conditions in Theorem 1 above (it is analytic in a vicinity of $\lambda = \lambda_0$ for all integers $m$, $n$ and all values of $v$ inside a certain domain, and the leading principal minors (\ref{eq14}) are nonzero in this domain). We also assume that the function $R_{n,m}(v,\lambda)$ is meromorphic in a vicinity of $\lambda = \lambda_0$ when $v$ takes values in the domain under consideration.
 
 Due to the Theorem 1 the discrete operator $L=D^{-1}_n PZ$ is reduced to the block-diagonal form $L_0 = D^{-1}_n h Z$ by the transformation $L \rightarrow T^{-1} L T = L_0$. It follows from (\ref{S3eq4}) that $[L_0, M_0] = 0$, where $M_0 := T^{-1} M T$. By the assumption above the coefficient $S$ in $M_0 = D^{-1}_m S$ is a formal series $S = (\lambda-\lambda_0)^k \sum_{i=0}^{\infty} S^{(i)} (\lambda-\lambda_0)^{-i}$.
 
\textbf{Theorem 2.} The coefficients $S^{(i)}$ of the series $S$ have the same block diagonal structure as the matrix $h$.

Scheme of the proof. The series $S$ and $h$ satisfy the equation $[D^{-1}_n h Z, D^{-1}_m S] = 0$ which implies
\begin{equation}	\label{S3eq5}
D_n(S) h = D_m(h)ZSZ^{-1}.
\end{equation} 
The last equation looks like (\ref{eq16}) with $S$ instead of $T$ and $D_m(h)$ instead of $P$. By using
the reasonings used in the proof of Theorem 1 one can check that the coefficients $S^{(i)}$ and $h^{(i)}$ of the series $S$ and $h$ have the same block diagonal structure. \fullsquare

Due to the block structure $S$ commutes with $Z$ and we find
\begin{equation}	\label{S3eq6}
D_n(S) h = D_m (h) S.
\end{equation}
By passing to the block entries $S = \{ S_{ij}\}$, $h = \{ h_{ij}\}$ in (\ref{S3eq6}) we get $D_n(S_{ii}) h_{ii} = D_m(h_{ii}) S_{ii}$. Now it is evident that the equation
\begin{equation}	\label{S3eq7}
(D_n - 1) \log \det S_{ii} = (D_m - 1) \log \det h_{ii}, \qquad i = 1,2,\ldots, r
\end{equation}
generates an infinite series of conservation laws for the equation (\ref{S3eq1}). Since the function $\det S = \prod_{i = 1}^{r} \det S_{ii}$ does not vanish identically then logarithms in (\ref{S3eq7}) are correctly defined.

\section{Conservation laws for the quad system connected with the series $A^{(1)}_{N}$ Kac-Moody Lie algebras}

In \cite{ALN} a finite field reduction is found for the discrete KP equation by Hirota \cite{Hirota2}
\begin{equation}\label{discreteKP}
\frac{pv^{(j-1)}_{1,0}v^{(j)}_{0,1}-qv^{(j-1)}_{0,1}v_{\vphantom{1}}^{(j)}}{v^{(j-1)}_{1,1}v_{\vphantom{1}}^{(j)}}=\frac{pv^{(j)}_{1,0}v^{(j+1)}_{0,1}-qv^{(j)}_{0,1}v^{(j+1)}}{v^{(j)}_{1,1}v_{\vphantom{1}}^{(j+1)}}
\end{equation}
where $j\in\{1,2,...,N'\}$ and $v^{(0)}=v^{(N'+1)}=1$, $p \neq 0$, $q \neq 0$.

For the case $N'=1$ the system coincides with the lattice mKdV equation \cite{Hirota2}. The hierarchies of conservation laws and symmetries to this equation are constructed in \cite{Xenitidis11}.
For $N'=2$ system (\ref{discreteKP}) corresponds to the lattice version of the modified Boussinesq equation \cite{npcq92}. Hierarchies of the symmetries and conservation laws of this model were described in \cite{xn12}.

It is remarkable that system (\ref{discreteKP}) is closely connected with the series $A^{(1)}_{N'}$ Kac-Moody Lie algebras. Recall that $A^{(1)}_{N'}$ can be  realized as the ring $sl(N'+1)$ of the $(N'+1)\times(N'+1)$ matrices with the vanishing trace. Introduce following  \cite{Dri} the system of the canonical generators $e_i$, $f_i$, $h_i$: $e_0=e_{1,N'+1}\lambda^{-1}$, $f_0=e_{N'+1,1}\lambda$, $h_0=e_{1,1}-e_{N'+1,N'+1}$, $e_i=e_{i+1,i}\lambda^{-1}$, $f_i=e_{i,i+1}\lambda$, $h_i=e_{i+1,i+1}-e_{i,i}$, $i=1,2,...,N'$.
Here $e_{i,j}$ is a matrix having  one  at the site $(i,j)$ and zeros elsewhere. 

Now the Lax pair for the system  (\ref{discreteKP}) found in \cite{ALN} can be rewritten as:
\begin{equation}  
y_{1,0} = f_py, \qquad y_{0,1} = f_qy, \label{S2eq2AN}
\end{equation}
where $f_p=-\bar \Lambda+ pe^{F_{1,0}-F}$, $f_q=-\bar \Lambda+ qe^{F_{0,1}-F}$ and $F=\sum_{i=1}^{N'} \log v^{(i)}h_i$, $\bar\Lambda=\sum_{i=0}^{N'} f_i$. The Lax pair (\ref{S2eq2AN}) is obtained from that given in \cite{ALN} by the following linear change of the variables $\tilde y(n)=\mathrm{diag} (1,v^{(1)},v^{(2)},...,v^{(N')})y(n)$. To the best of our knowledge the problem of assigning an integrable quad system with the related Lax pair to the affine Lie algebras is not completely solved. For more information on this issue see \cite{HabGarYang,HabGarArxiv,RalphWillox} and references therein.

The Lie algebraic language allows one to represent the potentials $f,g$ in an elegant form. However, in studying the asymptotics we have to leave the Lie algebra since the series $T$, $h$ belong to the corresponding loop groups.

\subsection{Lattice mKdV equation}

For $N'=1$ system (\ref{discreteKP}) corresponds to the quad equation
\begin{equation} \label{ALN1main}
\frac{p v_{1,0} - q v_{0,1}}{v} = \frac{p v_{0,1} - q v_{1,0}}{v_{1,1}}
\end{equation}
being the discrete version of the mKdV equation. It admits the Lax pair
\begin{equation}  
y_{1,0} = fy, \qquad y_{0,1} = gy, \label{ALN1LaxPair}
\end{equation}
where
\begin{equation*}
f = \left(  \begin{array}{cc}
p \frac{v_{1,0}}{v} & -\lambda\\
-\lambda & p \frac{v}{v_{1,0}}
\end{array} \right), \quad g = \left(  \begin{array}{cc}
q \frac{v_{0,1}}{v} & -\lambda\\
-\lambda & q \frac{v}{v_{0,1}}
\end{array} \right).
\end{equation*}
Since $\det f = p^2 - \lambda^2$ then the first equation in (\ref{ALN1LaxPair}) has three singular points: $\lambda = \infty$, $\lambda = \pm p$. Due to the Proposition (see \S 2) the singularity at $\lambda = \infty$ is removable. Concentrate on $\lambda_0 = p$. Let us reduce the equation to the desired special form (\ref{MainSpecialForm}).

It can be easily checked that the factorization (\ref{eq11}) for this case is
\begin{equation} \label{ALN1eq5}
f = \alpha Z \beta
\end{equation}
where
\begin{equation*}
\alpha = \left(  \begin{array}{cc}
p \frac{v_{1,0}}{v} & -1\\
-(p+\xi^{-1}) & -\frac{v}{v_{1,0}}
\end{array} \right), \quad \beta = \left(  \begin{array}{cc}
1 & -\frac{v}{v_{1,0}} \\
0 & 1
\end{array} \right), \quad Z = \left( \begin{array}{cc}
1 & 0\\
0 & \xi^{-1}
\end{array} \right).
\end{equation*}
Here $\lambda - p = \xi^{-1}$. Evidently the singular point $\lambda_0 = p$ is transformed to $\xi_0 = \infty$.

Remark that factorization (\ref{eq11}) is not unique. We have chosen (\ref{ALN1eq5}) such that the factor $\beta$ does not depend on $\xi$. Now to get the special form (\ref{MainSpecialForm}) change the variables $\psi = \beta y$:
\begin{equation}	\label{ALN1eq6}
\psi_{1,0} = P(\xi) Z \psi,
\end{equation}
where $P(\xi) = D_n(\beta) \alpha$:
\begin{equation}  \label{ALN1P}
P(\xi) = \left( \begin{array}{cc}
p \left( \frac{v_{1,0}}{v} + \frac{v_{1,0}}{v_{2,0}} \right) & -1 + \frac{v}{v_{2,0}}\\
-p & -\frac{v}{v_{1,0}}
\end{array} \right) + \left(  \begin{array}{cc}
\frac{v_{1,0}}{v_{2,0}} & 0\\
-1 & 0
\end{array} \right) \xi^{-1}.
\end{equation}
Function $P(\xi)$ satisfies the settings of the Theorem 1, indeed  the functions $\det P(\xi = \infty) = -2p $ and $\det_1 P(\xi = \infty) = p \left(  \frac{v_{1,0}}{v} + \frac{v_{1,0}}{v_{2,0}}\right) $ do not vanish if the variable $v = v(n,m)$ satisfies the inequalities $v_{2,0}v_{1,0} \neq - v_{1,0} v$, $v \neq 0$ for all $n$ and $m$. Therefore equation (\ref{ALN1eq6}) can be diagonalized, i.e. there exist formal series
\begin{equation}
T = T^{(0)} + T^{(1)} \xi^{-1} = \cdots, \quad h = h^{(0)} + h^{(1)} \xi ^{-1} + \cdots
\end{equation}
such that the formal change of the variables $\psi = T \varphi$ converts system (\ref{ALN1eq6}) to the system of the block-diagonal form
\begin{equation} \label{ALN1eq9}
\varphi_{1,0} = h Z \varphi.
\end{equation}
The coefficients $T^{(k)}$, $h^{(k)}$ are found by solving consecutively the set of equations (\ref{eq24}), (\ref{eq25}), (\ref{eq26}). Since the algorithm of solving these equations has been discussed earlier (see the proof of Theorem 1 above) we give only the answers:
\begin{eqnarray*}
T^{(0)} = \left( \begin{array}{cc}
1 & 0\\
-\frac{v_{-1,0}v_{1,0}}{v(v_{1,0} + v_{-1,0})} & 1
\end{array} \right), \\
T^{(1)} = \left( \begin{array}{cc}
0 & -\frac{v (v - v_{2,0})}{p v_{1,0}(v_{2,0}+v)}\\
\frac{v_{-1,0} v^2_{1,0}(v_{-2,0} - v)}{pv(v+ v_{-2,0})(v_{1,0}+v_{-1,0})^2} & 0
\end{array} \right),
\end{eqnarray*}
\begin{eqnarray*}
h^{(0)} = \left( \begin{array}{cc}
p \frac{v_{1,0}(v_{2,0} + v)}{v v_{2,0}} & 0\\
0 & -\frac{2 v v_{2,0}}{v_{1,0}(v_{2,0} + v)}
\end{array} \right),\\
h^{(1)} = \left( \begin{array}{cc}
\frac{v_{1,0}(v_{-1,0}v_{2,0} + v v_{1,0})}{v v_{2,0} (v_{1,0} + v_{-1,0})} & 0\\
0 & -\frac{v v_{2,0}(-3v v_{3,0} + v v_{1,0} + v_{2,0} v_{3,0} + v_{1,0}v_{2,0})}{p v_{1,0}(v_{3,0} + v_{1,0})(v + v_{2,0})^2}
\end{array} \right).
\end{eqnarray*}
According to the general scheme the second equation of the Lax pair (\ref{ALN1LaxPair}) is diagonalized by the same linear change of the variables. It takes the form
\begin{equation*}
\varphi_{0,1} = S \varphi,
\end{equation*}
where
\begin{eqnarray*}
S = S^{(0)} + S^{(1)} \xi^{-1} + \cdots,\\
S^{(0)} = \left( \begin{array}{cc}
(p^2 - q^2) \frac{v_{1,0} v_{0,1}}{v (v_{0,1} p - q v_{1,0})} & 0\\
0 & -\frac{v(v_{0,1}p - q v_{1,0})}{v_{0,1} v _{1,0}}
\end{array} \right),\\
S^{(1)} = \left( \begin{array}{cc}\frac{\bigl( pv_{0,1}(v_{-1,0} + v_{1,0}) - q (v_{-1,0} v_{1,0} + v^2_{0,1}) \bigr)v_{1,0}}{(v_{-1,0} + v_{1,0})(pv_{0,1} - qv_{1,0})v} & 0\\
0 & -\frac{2 v v_{2,0}}{v_{1,0}(v + v_{2,0})}
 \end{array} \right).
\end{eqnarray*}

Now the conservation laws can be derived from the relation
\begin{equation*}
(D_m - 1) \log h = (D_n - 1) \log S.
\end{equation*}
We write down in an explicit form three conservation laws from the infinite sequence obtained by the diagonalization procedure
\begin{eqnarray}
\fl (D_m - 1) \log \frac{v + v_{2,0}}{v_{2,0}} = (D_n - 1) \log \frac{v_{1,0} }{pv_{0,1} - q v_{1,0}}, \label{ALN1logCL}\\
\fl (D_m - 1) \frac{v_{-1,0}v_{2,0} + v v_{1,0}}{p(v_{-1,0} + v_{1,0})(v + v_{2,0})}  \nonumber\\
=(D_n - 1) \frac{pv_{0,1}(v_{-1,0} + v_{1,0}) - q(v_{-1,0} v_{1,0} + v^2_{0,1})}{(p^2-q^2)v_{0,1}(v_{-1,0}+v_{1,0})}, \label{ALN1CL2}\\
\fl (D_m - 1) \frac{1}{2} \frac{-3v v_{3,0} + v v_{1,0} + v_{2,0}v_{3,0} + v_{1,0}v_{2,0}}{p(v + v_{2,0})(v_{1,0} + v_{3,0})} 
= (D_n - 1) \frac{2 v_{0,1} v_{2,0}}{(v + v_{2,0})(p v_{0,1} - q v_{1,0})}. \label{ALN1CL3}
\end{eqnarray}
Since the equation (\ref{ALN1main}) is invariant under the replacing $n \leftrightarrow m$ one can easily find from the explicit formulas (\ref{ALN1logCL})-- (\ref{ALN1CL3}) conservation laws on the other direction.

As it is expected the conservation laws coincide with those found earlier in \cite{Xenitidis11}.

Let us give also expression of the first conservation law in terms of the matrix entries of the potentials $f$ and $g$:
\begin{equation}  \label{ALN1logCL2}
(D_m - 1) \log \det_1 P^{(0)} = (D_n - 1) \log \frac{1}{f_{22}- g_{22}},
\end{equation}
where $\det_1 P^{(0)}$ is the first order leading principal minor of the matrix $P^{(0)}$ defined in (\ref{ALN1P}).

\subsection{Lattice version of the modified Boussinesq equation }

For $N'=2$ we have the system
\begin{equation} \label{ALN2main}
\eqalign{\frac{p v^{(1)}_{1,0}- qv^{(1)}_{0,1}}{v_{\vphantom{1}}^{(1)}} = \frac{p v^{(1)}_{0,1} v^{(2)}_{1,0} - q v^{(1)}_{1,0} v^{(2)}_{0,1}}{v^{(1)}_{1,1} v_{\vphantom{1}}^{(2)}}, \cr  \frac{p v^{(1)}_{0,1} v^{(2)}_{1,0} - q v^{(1)}_{1,0} v^{(2)}_{0,1}}{v^{(1)}_{1,1}  v_{\vphantom{1}}^{(2)}} = \frac{p v^{(2)}_{0,1} - q v^{(2)}_{1,0}}{v^{(2)}_{1,1}}.}  
\end{equation}
and the Lax pair
\begin{equation}  
y_{1,0} = fy, \qquad y_{0,1} = gy, \label{ALN2LaxPair}
\end{equation}
where 
\begin{equation*}
f = \left(  \begin{array}{ccc}
p \frac{v^{(1)}_{1,0}}{v_{\vphantom{1}}^{(1)}} & -\lambda & 0\\
0 & p \frac{v_{\vphantom{1}}^{(1)}v^{(2)}_{1,0}}{v^{(1)}_{1,0}v_{\vphantom{1}}^{(2)}} & -\lambda\\
-\lambda & 0 & p \frac{v_{\vphantom{1}}^{(2)}}{v^{(2)}_{1,0}}
\end{array} \right), \qquad 
g = \left(  \begin{array}{ccc}
q \frac{v^{(1)}_{0,1}}{v_{\vphantom{1}}^{(1)}} & -\lambda & 0\\
0 & q \frac{v_{\vphantom{1}}^{(1)}v^{(2)}_{0,1}}{v^{(1)}_{0,1}v_{\vphantom{1}}^{(2)}} & -\lambda\\
-\lambda & 0 & q \frac{v_{\vphantom{1}}^{(2)}}{v^{(2)}_{0,1}}
\end{array} \right).
\end{equation*}
Since $\det f = p^3 - \lambda^3$ then the first system in (\ref{ALN2LaxPair}) has four singular points: $\lambda = \infty$, $\lambda = p$, $\lambda = pe^{\frac{2\pi i}{3}}$, $\lambda = pe^{\frac{-2\pi i}{3}}$. Due to the Proposition (see \S 2) the singularity at $\lambda = \infty$ is removable. Concentrate on $\lambda_0 = p$. Let us reduce the equation to the desired special form (\ref{MainSpecialForm}).

It can be easily checked that the factorization (\ref{eq11}) for this case is
\begin{equation} \label{ALN2eq15}
f = \alpha Z \beta
\end{equation}
where
\begin{eqnarray*}
\alpha = \left(  \begin{array}{ccc}
p \frac{v^{(1)}_{1,0}}{v_{\vphantom{1}}^{(1)}} & -(p+\xi^{-1}) & -\frac{v^{(1)}_{1,0} v_{\vphantom{1}}^{(2)}}{v_{\vphantom{1}}^{(1)} v^{(2)}_{1,0}}\\
0 & p \frac{v_{\vphantom{1}}^{(1)} v^{(2)}_{1,0}}{v^{(1)}_{1,0}v_{\vphantom{1}}^{(2)}} & -1\\
-(p+\xi^{-1}) & 0 & -\frac{v_{\vphantom{1}}^{(2)}}{v^{(2)}_{1,0}}
\end{array} \right), \\
\beta = \left(  \begin{array}{ccc}
1 & 0 & -\frac{v_{\vphantom{1}}^{(2)}}{v^{(2)}_{1,0}}\\
0 & 1 & -\frac{v^{(1)}_{1,0}v_{\vphantom{1}}^{(2)}}{v^{(1)} v^{(2)}_{1,0}}\\
0 & 0 & 1
\end{array} \right),
 \qquad Z = \left( \begin{array}{ccc}
1 & 0 & 0\\
0 & 1 & 0\\
0 & 0 & \xi^{-1}
\end{array} \right).
\end{eqnarray*}
Here $\lambda - p = \xi^{-1}$. The singular point $\lambda_0 = p$ is transformed to $\xi_0 = \infty$.

                               Now to get the special form (\ref{MainSpecialForm}) change the variables $\psi = \beta y$:
\begin{equation}	\label{ALN2eq16}
\psi_{1,0} = P(\xi) Z \psi,
\end{equation}
where $P(\xi) = D_n(\beta) \alpha$:
\begin{eqnarray} 
P(\xi) = P^{(0)} + P^{(1)} \xi^{-1}, \label{ALN2P}\\
P^{(0)} = \left( \begin{array}{ccc}
p \frac{v^{(1)}_{1,0}v^{(2)}_{2,0} + v_{\vphantom{1}}^{(1)} v^{(2)}_{1,0}}{v_{\vphantom{1}}^{(1)} v^{(2)}_{2,0}} & -p & \frac{v_{\vphantom{1}}^{(2)}\left(v_{\vphantom{1}}^{(1)} v^{(2)}_{1,0} - v^{(1)}_{1,0} v^{(2)}_{2,0}\right)}{v_{\vphantom{1}}^{(1)} v^{(2)}_{1,0} v^{(2)}_{2,0}}\\
p \frac{v^{(1)}_{2,0}v^{(2)}_{1,0}}{v^{(1)}_{1,0}v^{(2)}_{2,0}} & p \frac{v_{\vphantom{1}}^{(1)} v^{(2)}_{1,0}}{v^{(1)}_{1,0} v_{\vphantom{1}}^{(2)}} & \frac{v^{(1)}_{2,0} v_{\vphantom{1}}^{(2)} - v^{(1)}_{1,0}v^{(2)}_{2,0}}{v^{(1)}_{1,0}v^{(2)}_{2,0}}\\
-p & 0 & -\frac{v_{\vphantom{1}}^{(2)}}{v^{(2)}_{1,0}}
\end{array} \right), \nonumber\\
P^{(1)} = 
\left(  \begin{array}{ccc}
\frac{v^{(2)}_{1,0}}{v^{(2)}_{2,0}} & -1 & 0\\
\frac{v^{(1)}_{2,0} v^{(2)}_{1,0}}{v^{(1)}_{1,0} v^{(2)}_{2,0}} & 0 & 0\\
-1 & 0 & 0
\end{array} \right).\nonumber
\end{eqnarray}
Function $P(\xi)$ satisfies the settings of the Theorem 1, indeed $\det P(\xi = \infty) = -3p^2 \neq 0$ and $\det_2 P(\xi = \infty) = p^2 \frac{v^{(2)}_{1,0}\left(v^{(1)}_{1,0}v^{(2)}_{2,0} + v_{\vphantom{1}}^{(1)} v^{(2)}_{1,0} + v^{(1)}_{2,0} v_{\vphantom{1}}^{(2)}\right)}{v^{(1)}_{1,0}v_{\vphantom{1}}^{(2)} v^{(2)}_{2,0}} \neq 0$ if the variables $v^{(1)} = v^{(1)}(n,m)$ and $v^{(2)} = v^{(2)}(n,m)$ satisfy the inequalities $v^{(1)}_{1,0}v^{(2)}_{2,0} + v_{\vphantom{1}}^{(1)} v^{(2)}_{1,0} + v^{(1)}_{2,0} v_{\vphantom{1}}^{(2)} \neq 0$, $ v_{\vphantom{1}}^{(1)}\neq 0$ and $v_{\vphantom{1}}^{(2)} \neq 0$ for all $n$ and $m$. Therefore equation (\ref{ALN2eq16}) can be block-diagonalized, i.e. there exist formal series
\begin{equation}
T = T^{(0)} + T^{(1)} \xi^{-1} = \cdots, \quad h = h^{(0)} + h^{(1)} \xi ^{-1} + \cdots
\end{equation}
such that the formal change of the variables $\psi = T \varphi$ converts system (\ref{ALN2eq16}) to a block-diagonal system of the form
\begin{equation} \label{ALN2eq19}
\varphi_{1,0} = h Z \varphi.
\end{equation}
The coefficients $T^{(k)}$, $h^{(k)}$ are found by solving consecutively the set of equations (\ref{eq24}), (\ref{eq25}), (\ref{eq26}). Let us denote $\delta = v^{(1)}_{1,0}v^{(2)}_{2,0} + v_{\vphantom{1}}^{(1)} v^{(2)}_{1,0} + v^{(1)}_{2,0} v_{\vphantom{1}}^{(2)} $ and  write the first several terms of the series $T$ explicitly:
\begin{eqnarray*}
T^{(0)} = \left( \begin{array}{ccc}
1 & 0 & 0\\
0 & 1 & 0\\
-\frac{v^{(1)}_{-1,0}v^{(2)}_{1,0}}{\delta_{-1,0}} & -\frac{v_{\vphantom{1}}^{(1)} v^{(2)}_{-1,0} v^{(2)}_{1,0}}{v^{(2)} \delta_{-1,0}} & 1
\end{array} \right), \\
T^{(1)} = \left( \begin{array}{ccc}
0 & 0 & -\frac{v_{\vphantom{1}}^{(2)}\left(-2 v^{(1)}_{1,0}v^{(2)}_{2,0} + v^{(1)}_{2,0} v_{\vphantom{1}}^{(2)} + v_{\vphantom{1}}^{(1)} v^{(2)}_{1,0} \right)}{p v^{(2)}_{1,0}\delta }\\
0 & 0 & \frac{v^{(1)}_{1,0} v_{\vphantom{1}}^{(2)} \left( v^{(1)}_{1,0} v^{(2)}_{2,0} - 2 v^{(1)}_{2,0} v_{\vphantom{1}}^{(2)} + v_{\vphantom{1}}^{(1)} v^{(2)}_{1,0} \right)}{p v_{\vphantom{1}}^{(1)} v^{(2)}_{1,0} \delta}\\
T^{(1)}_{31} & T^{(1)}_{32} & 0
\end{array} \right),
\end{eqnarray*}
where
\begin{eqnarray*}
\fl T^{(1)}_{31} = \frac{3 v_{\vphantom{1}}^{(1)} v^{(1)}_{-1,0} v^{(2)}_{-1,0} v^{(2)}_{1,0} \left(v^{(1)}_{-2,0} v^{(2)}_{1,0} - v^{(1)}_{1,0}v^{(2)}_{-2,0}\right)}{p \delta^2_{-1,0} \delta_{-2,0}}  \\
+\frac{v^{(1)}_{-1,0} v^{(2)}_{1,0}\left(v_{\vphantom{1}}^{(1)} v^{(2)}_{1,0} - v^{(1)}_{1,0} v^{(2)}_{-1,0}\right)}{p \delta^2_{-1,0}},\\
\fl T^{(1)}_{32} = \frac{3 \left(v_{\vphantom{1}}^{(1)}\right)^2 v^{(2)}_{-1,0} v^{(2)}_{1,0}\left( v^{(1)}_{-2,0} v^{(2)}_{-1,0} v^{(2)}_{1,0} + v_{\vphantom{1}}^{(1)}  v^{(2)}_{-2,0} v^{(2)}_{1,0}\right) + v^{(1)}_{-1,0} v_{\vphantom{1}}^{(2)} v^{(2)}_{-2,0}}{pv^{(2)}\delta^2_{-1,0} \delta_{-2,0}} \\
- \frac{v_{\vphantom{1}}^{(1)}  v^{(2)}_{-1,0} v^{(2)}_{1,0}\left(2 v_{\vphantom{1}}^{(1)} v^{(2)}_{1,0} + v^{(1)}_{-1,0} v_{\vphantom{1}}^{(2)}\right)}{p v_{\vphantom{1}}^{(2)} \delta^2_{-1,0}}.
\end{eqnarray*}
Let us write the first terms of the series $h$ explicitly:
\begin{eqnarray*}
h^{(0)} = \left( \begin{array}{ccc}
p\frac{v^{(1)}_{1,0}v^{(2)}_{2,0} + v_{\vphantom{1}}^{(1)} v^{(2)}_{1,0}}{v_{\vphantom{1}}^{(1)} v^{(2)}_{2,0}} & -p & 0\\
p \frac{v^{(1)}_{2,0}v^{(2)}_{1,0}}{v^{(1)}_{1,0} v^{(2)}_{2,0}} & p \frac{v_{\vphantom{1}}^{(1)} v^{(2)}_{1,0}}{v^{(1)}_{1,0} v_{\vphantom{1}}^{(2)}} & 0\\
0 & 0 & \frac{3 v^{(1)}_{1,0}v_{\vphantom{1}}^{(2)} v^{(2)}_{2,0}}{v^{(2)}_{1,0} \delta}
\end{array} \right),\\
h^{(1)} = \left( \begin{array}{ccc}
h^{(1)}_{11} & h^{(1)}_{12} & 0\\
h^{(1)}_{21} & h^{(1)}_{22} & 0\\
0 & 0 & h^{(1)}_{33}
\end{array} \right),
\end{eqnarray*}
where
\begin{eqnarray*}
h^{(1)}_{11} = -\frac{v^{(1)}_{-1,0} v_{\vphantom{1}}^{(2)} \left(v_{\vphantom{1}}^{(1)} v^{(2)}_{1,0} - v^{(1)}_{1,0} v^{(2)}_{2,0}\right) }{v_{\vphantom{1}}^{(1)} v^{(1)}_{2,0} \delta_{-1,0}} + \frac{v^{(2)}_{1,0}}{v^{(2)}_{2,0}},\\
h^{(1)}_{12} = -1- \frac{v_{\vphantom{1}}^{(2)} v^{(2)}_{-1,0}\left(v_{\vphantom{1}}^{(1)} v^{(2)}_{1,0} - v^{(1)}_{1,0} v^{(2)}_{2,0}\right)}{v_{\vphantom{1}}^{(2)}v^{(2)}_{2,0} \delta_{-1,0}},\\
h^{(1)}_{21} = \frac{v^{(1)}_{-1,0} v^{(2)}_{1,0} \left(v^{(1)}_{1,0} v^{(2)}_{2,0} - v_{\vphantom{1}}^{(2)}v^{(1)}_{2,0}\right)}{v^{(1)}_{1,0} v^{(2)}_{2,0} \delta_{-1,0}},\\
h^{(1)}_{22} = \frac{v_{\vphantom{1}}^{(1)} v^{(2)}_{-1,0} v^{(2)}_{1,0} \left(v^{(1)}_{1,0} v^{(2)}_{2,0} - v^{(1)}_{2,0} v_{\vphantom{1}}^{(2)}\right)}{v^{(1)}_{1,0} v_{\vphantom{1}}^{(2)} v^{(2)}_{2,0} \delta_{-1,0}},\\
\fl h^{(1)}_{33} = \frac{3 v^{(1)}_{1,0} v^{(1)}_{2,0} \left(v_{\vphantom{1}}^{(2)}\right)^2 v^{(2)}_{2,0} \left(v^{(1)}_{2,0} v^{(2)}_{3,0} - 2 v^{(1)}_{3,0}v^{(2)}_{1,0} + v^{(1)}_{1,0}v^{(2)}_{2,0}  \right)}{p v^{(2)}_{1,0} \delta \delta^2_{1,0} } \\
+ \frac{3 v^{(1)}_{1,0} v_{\vphantom{1}}^{(2)} v^{(2)}_{1,0} v^{(2)}_{2,0} \left(v^{(1)}_{2,0} v_{\vphantom{1}}^{(2)} - v^{(1)}_{1,0} v^{(2)}_{2,0}\right)}{p \left(v^{(2)}_{1,0}\right)^2 \delta^2}.      
\end{eqnarray*} 

According to the general scheme the second equation of the Lax pair (\ref{ALN2LaxPair}) is block-diagonalized by the same linear change of the variables. It takes the form
\begin{equation*}
\varphi_{0,1} = S \varphi,
\end{equation*}
where
\begin{eqnarray*}
S = S^{(0)} + S^{(1)} \xi^{-1} + \cdots,\\
S^{(0)} =  \left( \begin{array}{ccc}
\frac{p^2 v^{(1)}_{1, 0} v^{(2)}_{0, 1} -q^2 v^{(1)}_{0, 1} v^{(2)}_{1, 0}}{\left(p v^{(2)}_{0, 1}-q v^{(2)}_{1, 0}\right) v^{(1)} } & -p & 0\\
p\frac{ v^{(2)}_{0, 1} \left(p v^{(1)}_{0, 1} v^{(2)}_{1, 0}-q v^{(1)}_{1, 0} v^{(2)}_{0, 1} \right)}{  \left(p v^{(2)}_{0, 1}-q v^{(2)}_{1, 0}\right) v^{(1)}_{0, 1}v_{\vphantom{1}}^{(2)}} & q\frac{ v_{\vphantom{1}}^{(1)}  v^{(2)}_{0, 1}}{v^{(1)}_{0, 1} v_{\vphantom{1}}^{(2)} } & 0\\
0 & 0 & -\frac{v_{\vphantom{1}}^{(2)}  \left(p v^{(2)}_{0, 1}-q v^{(2)}_{1, 0}\right)}{v^{(2)}_{1, 0} v^{(2)}_{0, 1}}
\end{array} \right),\\
S^{(1)} = \left(  \begin{array}{ccc}
S^{(1)}_{11} & S^{(1)}_{12} & 0\\
S^{(1)}_{21} & S^{(1)}_{22} & 0\\
0 & 0 & S^{(1)}_{33}
\end{array} \right),
\end{eqnarray*}
where
\begin{eqnarray*}
S^{(1)}_{11} = \frac{q v^{(1)}_{-1, 0} v_{\vphantom{1}}^{(2)}  \left(v^{(1)}_{0, 1} v^{(2)}_{0, 1}-v^{(1)}_{1, 0} v^{(2)}_{1, 0}\right) }{v_{\vphantom{1}}^{(1)}  \left(p v^{(2)}_{0, 1}-q v^{(2)}_{1, 0}\right) \delta_{-1,0}} + \frac{v^{(2)}_{0, 1} \left(v^{(1)}_{1, 0} p-q v^{(1)}_{0, 1}\right)}{v_{\vphantom{1}}^{(1)}  \left(p v^{(2)}_{0, 1}-q v^{(2)}_{1, 0}\right)},\\
S^{(1)}_{12} = -1 + \frac{q \left(v^{(1)}_{0, 1} v^{(2)}_{0, 1}-v^{(1)}_{1, 0} v^{(2)}_{1, 0}\right) v^{(2)}_{-1, 0}}{\left(p v^{(2)}_{0, 1}-q v^{(2)}_{1, 0}\right) \delta_{-1,0}},\\
S^{(1)}_{21} = \frac{q \left(v^{(1)}_{0, 1} \left(v^{(2)}_{1, 0}\right)^2-\left(v^{(2)}_{0, 1}\right)^2 v^{(1)}_{1, 0}\right) v^{(1)}_{-1, 0}}{v^{(1)}_{0, 1} \left(p v^{(2)}_{0, 1}-q v^{(2)}_{1, 0}\right) \delta_{-1,0}} + \frac{v^{(2)}_{0, 1} \left(p v^{(1)}_{0, 1}  v^{(2)}_{1, 0}-q v^{(2)}_{0, 1} v^{(1)}_{1, 0}\right)}{v_{\vphantom{1}}^{(2)}  \left(p v^{(2)}_{0, 1}-q v^{(2)}_{1, 0}\right) v^{(1)}_{0, 1}},\\
S^{(1)}_{22} = \frac{q \left(v^{(1)}_{0, 1} \left(v^{(2)}_{1, 0}\right)^2-\left(v^{(2)}_{0, 1}\right)^2 v^{(1)}_{1, 0}\right) v_{\vphantom{1}}^{(1)}  v^{(2)}_{-1, 0}}{v_{\vphantom{1}}^{(2)}  \left(p v^{(2)}_{0, 1}-q v^{(2)}_{1, 0}\right) v^{(1)}_{0, 1} \delta_{-1,0}},\\
S^{(1)}_{33} = \frac{3 v^{(2)}_{2, 0} v^{(1)}_{1, 0} v^{(2)}_{0, 1}}{\delta \left(p v^{(2)}_{0, 1}-q v^{(2)}_{1, 0}\right)}.
\end{eqnarray*}

Now let us write down in an explicit form three conservation laws from the infinite sequence obtained by the diagonalization procedure
\begin{eqnarray}
\fl (D_m - 1) \log \frac{v^{(1)}_{1,0}v^{(2)}_{2,0} + v_{\vphantom{1}}^{(1)} v^{(2)}_{1,0} + v^{(1)}_{2,0} v_{\vphantom{1}}^{(2)} }{v^{(1)}_{1,0} v^{(2)}_{2,0}}  = (D_n - 1) \log \frac{v^{(2)}_{1,0} }{ p v^{(2)}_{0,1} - q v^{(2)}_{1,0}}, \label{ALN2logCL}\\
\fl (D_m -1) \left[\frac{v^{(1)}_{1,0}v^{(2)}_{-1,0} \left(v^{(1)}_{1,0}v^{(2)}_{2,0} - 2 v^{(1)}_{2,0}v_{\vphantom{1}}^{(2)} + v_{\vphantom{1}}^{(1)} v^{(2)}_{1,0} \right) }{p \delta \delta_{-1,0}}  \right.\nonumber\\
\left.-\frac{v^{(1)}_{-1,0}v_{\vphantom{1}}^{(2)}\left(-2 v^{(1)}_{1,0}v^{(2)}_{2,0} + v^{(1)}_{2,0}v_{\vphantom{1}}^{(2)} + v_{\vphantom{1}}^{(1)} v^{(2)}_{1,0}\right)}{p \delta \delta_{-1,0}} + \frac{2 v^{(1)}_{2,0}v_{\vphantom{1}}^{(2)} + v_{\vphantom{1}}^{(1)} v^{(2)}_{1,0}}{p \delta} \right]\nonumber\\
\fl =(D_n -1)\left[ \frac{q v^{(1)}_{-1,0}v_{\vphantom{1}}^{(2)} \left( p \left( v^{(1)}_{1,0} \left(v^{(2)}_{0,1}\right)^2  - v^{(1)}_{0,1} \left(v^{(2)}_{1,0}\right)^2\right) + q v^{(2)}_{0,1} \left( v^{(1)}_{0,1} v^{(2)}_{0,1} - v^{(1)}_{1,0}v^{(2)}_{1,0} \right)\right)  }{(p^3-q^3)v^{(1)}_{0,1}v^{(2)}_{0,1}v^{(2)}_{1,0}\delta_{-1,0}} \right.\nonumber \\
- \frac{q v^{(2)}_{-1,0}\left( p v^{(2)}_{0,1} \left( \left(v^{(1)}_{0,1}\right)^2 v^{(2)}_{1,0} - \left(v^{(1)}_{1,0}\right)^2v^{(2)}_{0,1} \right) + q v^{(1)}_{0,1} \left( v^{(1)}_{0,1} \left(v^{(2)}_{1,0}\right)^2 - v^{(1)}_{1,0}\left(v^{(2)}_{0,1}\right)^2 \right)\right)}{(p^3-q^3)v^{(1)}_{0,1}v^{(2)}_{0,1}v^{(2)}_{1,0}\delta_{-1,0}} \nonumber\\
\left.+\frac{2p^2 v^{(1)}_{0,1} v^{(2)}_{1,0} - pq v^{(1)}_{1,0}v^{(2)}_{0,1} - q^2 v^{(1)}_{0,1}v^{(2)}_{0,1}}{(p^3-q^3)v^{(1)}_{0,1}v^{(2)}_{1,0}}\right], \label{ALN2CL2}\\
\fl (D_m - 1) \left[\frac{v_{\vphantom{1}}^{(1)} v^{(2)}_{1,0} \left(-2 v^{(1)}_{2,0}v^{(2)}_{3,0} + v^{(1)}_{3,0}v^{(2)}_{1,0} + v^{(1)}_{1,0} v^{(2)}_{2,0}\right)}{p \delta \delta_{1,0}} \right.\nonumber\\
\left.-\frac{v^{(2)}v^{(1)}_{2,0}\left(v^{(1)}_{2,0}v^{(2)}_{3,0} -2 v^{(1)}_{3,0} v^{(2)}_{1,0} + v^{(1)}_{1,0}v^{(2)}_{2,0}\right)}{p \delta \delta_{1,0}} + \frac{v^{(1)}_{1,0}v^{(2)}_{2,0}-v^{(1)}_{2,0}v_{\vphantom{1}}^{(2)}}{p \delta} \right]\nonumber\\
=(D_n -1) \frac{3 v^{(1)}_{1,0} v^{(2)}_{0,1} v^{(2)}_{2,0}}{\delta \left(p v^{(2)}_{0,1} - q v^{(2)}_{1,0}\right)}.\label{ALN2CL3}
\end{eqnarray}
Recall that $\delta =v^{(1)}_{1,0}v^{(2)}_{2,0} + v_{\vphantom{1}}^{(1)} v^{(2)}_{1,0} + v^{(1)}_{2,0} v_{\vphantom{1}}^{(2)} $.

Since the equation (\ref{ALN2main}) is invariant under the replacing $n \leftrightarrow m$ one can easily find from the explicit formulas (\ref{ALN2logCL})-- (\ref{ALN2CL3}) conservation laws on the other direction.

The conservation laws (\ref{ALN2logCL}), (\ref{ALN2CL2}) are equivalent to those given earlier in \cite{xn12} (up to a misprint in \cite{xn12}).

Let us give also expression of the first conservation law in terms of the matrix entries of the potentials $f$ and $g$:
\begin{equation}  \label{ALN2logCL2}
(D_m - 1) \log \det_2 P^{(0)} = (D_n - 1) \log \frac{1}{f_{33}- g_{33}},
\end{equation}
where $\det_2 P^{(0)}$ is the second order leading principal minor of the matrix $P^{(0)}$ defined in (\ref{ALN2P}).

\subsection{The system connected with the series $A^{(1)}_{3}$}

For $N'=3$ we have the system
\begin{equation} \label{ALN3main}
\eqalign{\frac{p v^{(1)}_{1,0}- qv^{(1)}_{0,1}}{v_{\vphantom{1}}^{(1)}} = \frac{p v^{(1)}_{0,1} v^{(2)}_{1,0} - q v^{(1)}_{1,0} v^{(2)}_{0,1}}{v^{(1)}_{1,1} v_{\vphantom{1}}^{(2)}}, \cr  \frac{p v^{(1)}_{0,1} v^{(2)}_{1,0} - q v^{(1)}_{1,0} v^{(2)}_{0,1}}{v^{(1)}_{1,1}  v_{\vphantom{1}}^{(2)}} = \frac{p v^{(2)}_{0,1} v^{(3)}_{1,0} -q v^{(2)}_{1,0} v^{(3)}_{0,1} }{v^{(2)}_{1,1} v_{\vphantom{1}}^{(3)}}, \cr \frac{p v^{(2)}_{0,1}v^{(3)}_{1,0} - q v^{(2)}_{1,0} v^{(3)}_{0,1}}{v^{(2)}_{1,1} v_{\vphantom{1}}^{(3)}} = \frac{p v^{(3)}_{0,1} - q v^{(3)}_{1,0}}{v^{(3)}_{1,1}}}
\end{equation}
and the Lax pair
\begin{equation}  
y_{1,0} = fy, \qquad y_{0,1} = gy, \label{ALN3LaxPair}
\end{equation}
where 
\begin{eqnarray*}
f = \left( \begin{array}{cccc}
p\frac{v^{(1)}_{1, 0}}{v_{\vphantom{1}}^{(1)} } & -\lambda & 0 & 0\\
0 & p \frac{v_{\vphantom{1}}^{(1)} v^{(2)}_{1, 0}  }{v^{(1)}_{1, 0} v_{\vphantom{1}}^{(2)}  } & -\lambda & 0 \\
0 & 0 & p\frac{v_{\vphantom{1}}^{(2)} v^{(3)}_{1, 0}  }{  v^{(2)}_{1, 0}v_{\vphantom{1}}^{(3)}} & -\lambda\\
-\lambda & 0 & 0 & p \frac{v_{\vphantom{1}}^{(3)} }{v^{(3)}_{1, 0}}
\end{array} \right), \\
g = \left( \begin{array}{cccc}
q\frac{v^{(1)}_{0, 1}}{v_{\vphantom{1}}^{(1)} } & -\lambda & 0 & 0\\
0 & q \frac{v_{\vphantom{1}}^{(1)} v^{(2)}_{0, 1}  }{ v^{(1)}_{0, 1} v_{\vphantom{1}}^{(2)} } & -\lambda & 0 \\
0 & 0 & q\frac{v_{\vphantom{1}}^{(2)} v^{(3)}_{0, 1}  }{v^{(2)}_{0, 1} v_{\vphantom{1}}^{(3)}  } & -\lambda\\
-\lambda & 0 & 0 & q \frac{v_{\vphantom{1}}^{(3)} }{v^{(3)}_{0, 1}}
\end{array} \right).
\end{eqnarray*}
Since $\det f = p^4 - \lambda^4$ the first system in (\ref{ALN3LaxPair}) has five singular points: $\lambda = \infty$, $\lambda = \pm p$,  $\lambda = \pm ip$. Due to the Proposition (see \S 2) the singularity at $\lambda = \infty$ is removable. Concentrate on $\lambda_0 = p$, the others are studied in a similar way. Let us reduce the equation to the desired special form (\ref{MainSpecialForm}). The factorization (\ref{eq11}) for this case is
\begin{equation} \label{ALN3eq25}
f = \alpha Z \beta
\end{equation}
where
\begin{eqnarray*}
\alpha = \left( \begin{array}{cccc}
p \frac{v^{(1)}_{1, 0}}{v^{(1)} } & -\lambda & 0 & -\frac{v^{(1)}_{1, 0} v_{\vphantom{1}}^{(3)} }{v_{\vphantom{1}}^{(1)}  v^{(3)}_{1, 0}}\\
0 & p \frac{v_{\vphantom{1}}^{(1)} v^{(2)}_{1, 0}  }{ v^{(1)}_{1, 0} v_{\vphantom{1}}^{(2)} } & -\lambda & -\frac{v^{(2)}_{1, 0} v_{\vphantom{1}}^{(3)}  }{v_{\vphantom{1}}^{(2)} v^{(3)}_{1, 0}  }\\
0 & 0 & p \frac{v_{\vphantom{1}}^{(2)} v^{(3)}_{1, 0}  }{ v^{(2)}_{1, 0} v_{\vphantom{1}}^{(3)} } & -1\\
-\lambda & 0 & 0 & -\frac{v_{\vphantom{1}}^{(3)} }{v^{(3)}_{1, 0}}
\end{array} \right), \\
\beta = \left(\begin{array}{cccc}
1 & 0 & 0 & -2 \frac{v_{\vphantom{1}}^{(3)} }{v^{(3)}_{1, 0}}\\
0 & 1 & 0 & -2 \frac{v^{(1)}_{1, 0} v_{\vphantom{1}}^{(3)} }{v_{\vphantom{1}}^{(1)}  v^{(3)}_{1, 0}}\\
0 & 0 & 1 & -2\frac{v^{(2)}_{1, 0}v_{\vphantom{1}}^{(3)}  }{v_{\vphantom{1}}^{(2)}v^{(3)}_{1, 0}  }\\
0 & 0 & 0 & 1
\end{array} \right),
 \qquad Z = \left( \begin{array}{cccc}
1 & 0 & 0 & 0\\
0 & 1 & 0 & 0 \\
0 & 0 & 1 & 0\\
0 & 0 & 0 & \xi^{-1}
\end{array} \right).
\end{eqnarray*}
Here $\lambda - p = \xi^{-1}$. The singular point $\lambda_0 = p$ is transformed to $\xi_0 = \infty$.

Now to get the special form (\ref{MainSpecialForm}) change the variables $\psi = \beta y$:
\begin{equation}	\label{ALN3eq26}
\psi_{1,0} = P(\xi) Z \psi,
\end{equation}
where $P(\xi) = D_n(\beta) \alpha$:
\begin{eqnarray} 
P(\xi) = P^{(0)} + P^{(1)} \xi^{-1}, \label{ALN3P}\\
P^{(0)} = \left( \begin{array}{cccc}
p \frac{ \left(v^{(1)}_{1, 0} v^{(3)}_{2, 0}+v_{\vphantom{1}}^{(1)}  v^{(3)}_{1, 0}\right)}{v_{\vphantom{1}}^{(1)}  v^{(3)}_{2, 0}} & -p & 0 & \frac{v_{\vphantom{1}}^{(3)}  \left(v_{\vphantom{1}}^{(1)}  v^{(3)}_{1, 0}-v^{(1)}_{1, 0} v^{(3)}_{2, 0}\right)}{v_{\vphantom{1}}^{(1)}  v^{(3)}_{1, 0} v^{(3)}_{2, 0}}\\
p \frac{v^{(1)}_{2, 0} v^{(3)}_{1, 0}}{v^{(1)}_{1, 0} v^{(3)}_{2, 0}} & p \frac{v_{\vphantom{1}}^{(1)} v^{(2)}_{1, 0}  }{ v^{(1)}_{1, 0} v_{\vphantom{1}}^{(2)} } & -p & -\frac{v_{\vphantom{1}}^{(3)}  \left(v^{(1)}_{1, 0}v^{(2)}_{1, 0}  v^{(3)}_{2, 0}-v^{(1)}_{2, 0} v_{\vphantom{1}}^{(2)} v^{(3)}_{1, 0}  \right)}{v^{(1)}_{1, 0} v_{\vphantom{1}}^{(2)} v^{(3)}_{1, 0}   v^{(3)}_{2, 0}}\\
p \frac{v^{(2)}_{2, 0} v^{(3)}_{1, 0} }{v^{(2)}_{1, 0} v^{(3)}_{2, 0} } & 0 & p\frac{ v_{\vphantom{1}}^{(2)} v^{(3)}_{1, 0} }{ v^{(2)}_{1, 0} v_{\vphantom{1}}^{(3)} } & -\frac{v^{(2)}_{1, 0} v^{(3)}_{2, 0} -v^{(2)}_{2, 0} v_{\vphantom{1}}^{(3)} }{v^{(2)}_{1, 0} v^{(3)}_{2, 0} }\\
-p & 0 & 0 & -\frac{v_{\vphantom{1}}^{(3)} }{v^{(3)}_{1, 0}}
\end{array} \right) \nonumber\\
P^{(1)} = 
\left(  \begin{array}{cccc}
\frac{v^{(3)}_{1, 0}}{v^{(3)}_{2, 0}} & -1 & 0 & 0\\
\frac{v^{(1)}_{2, 0} v^{(3)}_{1, 0}}{v^{(1)}_{1, 0} v^{(3)}_{2, 0}} & 0 & -1 & 0\\
\frac{v^{(2)}_{2, 0} v^{(3)}_{1, 0} }{v^{(2)}_{1, 0} v^{(3)}_{2, 0} } & 0 & 0 & 0\\
-1 & 0 & 0 & 0
\end{array} \right).\nonumber
\end{eqnarray}
Function $P(\xi)$ satisfies the settings of the Theorem 1, indeed $\det P(\xi = \infty) = -4 p^3 \neq 0$ and $\det_3 P(\xi = \infty) =p^3\frac{v^{(3)}_{1, 0} \left(v^{(1)}_{1, 0} v^{(2)}_{2, 0} v_{\vphantom{1}}^{(3)}  +v^{(1)}_{2, 0} v_{\vphantom{1}}^{(2)} v^{(3)}_{1, 0}  +v^{(1)}_{1, 0} v^{(2)}_{1, 0}  v^{(3)}_{2, 0}+v_{\vphantom{1}}^{(1)} v^{(2)}_{1, 0}   v^{(3)}_{1, 0}\right)}{  v^{(1)}_{1, 0} v^{(2)}_{1, 0}  v_{\vphantom{1}}^{(3)} v^{(3)}_{2, 0}} \neq 0$ if the variables $v^{(1)} = v^{(1)}(n,m)$, $v^{(2)} = v^{(2)}(n,m)$ and $v^{(3)} = v^{(3)}(n,m)$ satisfy the inequalities \\
$v^{(1)}_{1, 0} v^{(2)}_{2, 0} v_{\vphantom{1}}^{(3)}  +v^{(1)}_{2, 0} v_{\vphantom{1}}^{(2)} v^{(3)}_{1, 0}  +v^{(1)}_{1, 0} v^{(2)}_{1, 0}  v^{(3)}_{2, 0}+v_{\vphantom{1}}^{(1)} v^{(2)}_{1, 0}   v^{(3)}_{1, 0} \neq 0$, 
 $ v^{(1)} v^{(2)} v^{(3)}\neq 0$ for all $n$ and $m$. Therefore equation (\ref{ALN3eq26}) can be block-diagonalized, i.e. there exist formal series
\begin{equation}
T = T^{(0)} + T^{(1)} \xi^{-1} = \cdots, \quad h = h^{(0)} + h^{(1)} \xi ^{-1} + \cdots
\end{equation}
such that the formal change of the variables $\psi = T \varphi$ converts system (\ref{ALN3eq26}) to the block-diagonal system of the form
\begin{equation} \label{ALN3eq}
\varphi_{1,0} = h Z \varphi.
\end{equation}
The coefficients $T^{(k)}$, $h^{(k)}$ are found by solving consecutively the set equations (\ref{eq24}), (\ref{eq25}), (\ref{eq26}). Let us denote $\delta = v^{(1)}_{1, 0} v^{(2)}_{2, 0} v_{\vphantom{1}}^{(3)}  +v^{(1)}_{2, 0} v_{\vphantom{1}}^{(2)} v^{(3)}_{1, 0}  +v^{(1)}_{1, 0} v^{(2)}_{1, 0}  v^{(3)}_{2, 0}+v^{(1)}_{\vphantom{1}} v^{(2)}_{1, 0}   v^{(3)}_{1, 0} $ and  write the first terms of the series $T$ and $h$ explicitly:
\begin{eqnarray*}
T^{(0)} = \left(  \begin{array}{cccc}
1 & 0 & 0 & 0\\
0 & 1 & 0 & 0\\
0 & 0 & 1 & 0\\
 -\frac{v^{(1)}_{-1, 0} v_{\vphantom{1}}^{(2)}  v^{(3)}_{1, 0} }{\delta_{-1,0}}  & -\frac{v_{\vphantom{1}}^{(1)} v^{(2)}_{-1, 0} v^{(3)}_{1, 0} }{\delta_{-1,0}} & -\frac{v_{\vphantom{1}}^{(1)} v_{\vphantom{1}}^{(2)} v^{(3)}_{-1, 0}   v^{(3)}_{1, 0} }{v_{\vphantom{1}}^{(3)}  \delta_{-1,0}} & 1
\end{array} \right),\\
h^{(0)} = \left(  \begin{array}{cccc}
p \frac{v^{(1)}_{1, 0} v^{(3)}_{2, 0}+v_{\vphantom{1}}^{(1)}  v^{(3)}_{1, 0}}{v_{\vphantom{1}}^{(1)}  v^{(3)}_{2, 0}} & -p & 0 & 0\\
p \frac{v^{(1)}_{2, 0} v^{(3)}_{1, 0}}{v^{(1)}_{1, 0} v^{(3)}_{2, 0}} & p\frac{v_{\vphantom{1}}^{(1)} v^{(2)}_{1, 0}  }{v^{(1)}_{1, 0} v_{\vphantom{1}}^{(2)}  } & -p & 0\\
p \frac{v^{(2)}_{2, 0} v^{(3)}_{1, 0} }{v^{(2)}_{1, 0} v^{(3)}_{2, 0} } & 0 & p\frac{v_{\vphantom{1}}^{(2)}  v^{(3)}_{1, 0}}{v^{(2)}_{1, 0} v_{\vphantom{1}}^{(3)}  } & 0\\
0 & 0 & 0 & -4\frac{v^{(1)}_{1, 0} v^{(2)}_{1, 0} v_{\vphantom{1}}^{(3)} v^{(3)}_{2, 0}  }{v^{(3)}_{1, 0} \delta}
\end{array} \right).
\end{eqnarray*}

According to the general scheme the second equation of the Lax pair (\ref{ALN3LaxPair}) is reduced to the block-diagonal form by the same linear change of the variables. It takes the form
\begin{equation*}
\varphi_{0,1} = S \varphi,
\end{equation*}
where
\begin{eqnarray*}
S = S^{(0)} + S^{(1)} \xi^{-1} + \cdots,\\
S^{(0)} = \left(  \begin{array}{cccc}
\frac{p^2 v^{(1)}_{1, 0} v^{(3)}_{0, 1}   -q^2 v^{(1)}_{0, 1} v^{(3)}_{1, 0}}{\left(p v^{(3)}_{0, 1}-q v^{(3)}_{1, 0}\right) v_{\vphantom{1}}^{(1)} } & -p & 0 & 0\\
p \frac{v^{(3)}_{0, 1} \left(p  v^{(1)}_{0, 1} v^{(2)}_{1, 0}-q v^{(1)}_{1, 0} v^{(2)}_{0, 1} \right)}{\left(p v^{(3)}_{0, 1}-q v^{(3)}_{1, 0}\right)v^{(1)}_{0, 1} v_{\vphantom{1}}^{(2)}  } & q \frac{v_{\vphantom{1}}^{(1)}  v^{(2)}_{0, 1}}{v^{(1)}_{0, 1} v_{\vphantom{1}}^{(2)} } & -p & 0\\ 
p \frac{v^{(3)}_{0, 1}\left(p v^{(2)}_{0, 1} v^{(3)}_{1, 0}-q v^{(2)}_{1, 0} v^{(3)}_{0, 1} \right)}{v_{\vphantom{1}}^{(3)}  \left(p v^{(3)}_{0, 1}-q v^{(3)}_{1, 0}\right) v^{(2)}_{0, 1}} & 0 & q\frac{v_{\vphantom{1}}^{(2)}  v^{(3)}_{0, 1}}{v^{(2)}_{0, 1} v_{\vphantom{1}}^{(3)} } & 0\\
0& 0 & 0 & -\frac{v_{\vphantom{1}}^{(3)}  \left(p v^{(3)}_{0, 1}-q v^{(3)}_{1, 0}\right)}{v^{(3)}_{1, 0} v^{(3)}_{0, 1}}
\end{array} \right).
\end{eqnarray*}

Now let us write down in an explicit form three conservation laws from the infinite sequence obtained by the diagonalization procedure
\begin{eqnarray}
\fl (D_m - 1) \log \frac{v^{(3)}_{1, 0} \left(v^{(1)}_{1, 0} v^{(2)}_{2, 0} v_{\vphantom{1}}^{(3)}  +v^{(1)}_{2, 0} v_{\vphantom{1}}^{(2)} v^{(3)}_{1, 0}  +v^{(1)}_{1, 0} v^{(2)}_{1, 0}  v^{(3)}_{2, 0}+v_{\vphantom{1}}^{(1)} v^{(2)}_{1, 0}   v^{(3)}_{1, 0}\right)}{ v^{(1)}_{1, 0}  v^{(2)}_{1, 0} v_{\vphantom{1}}^{(3)} v^{(3)}_{2, 0}} \nonumber\\
 =(D_n - 1) \log \frac{v^{(3)}_{1, 0} v^{(3)}_{0, 1}}{v_{\vphantom{1}}^{(3)}  \left(p v^{(3)}_{0, 1}-q v^{(3)}_{1, 0}\right)}, \label{ALN3logCL}
\end{eqnarray}

\begin{eqnarray}
\fl (D_m - 1)\left[  -\frac{  \left(-3 v^{(1)}_{1, 0} v^{(2)}_{1, 0}  v^{(3)}_{2, 0}+v^{(1)}_{1, 0} v^{(2)}_{2, 0} v_{\vphantom{1}}^{(3)}  +v^{(1)}_{2, 0} v_{\vphantom{1}}^{(2)} v^{(3)}_{1, 0}  + v_{\vphantom{1}}^{(1)} v^{(2)}_{1, 0}  v^{(3)}_{1, 0}\right) v^{(1)}_{-1, 0} v_{\vphantom{1}}^{(2)} v_{\vphantom{1}}^{(3)} }{p\delta \delta_{-1, 0}  } \right.\nonumber \\
+\frac{2   \left(v^{(1)}_{1, 0} v^{(2)}_{1, 0}  v^{(3)}_{2, 0}-v^{(1)}_{1, 0}v^{(2)}_{2, 0} v_{\vphantom{1}}^{(3)}  + v_{\vphantom{1}}^{(1)} v^{(2)}_{1, 0}  v^{(3)}_{1, 0}-v^{(1)}_{2, 0} v_{\vphantom{1}}^{(2)} v^{(3)}_{1, 0}  \right) v^{(1)}_{1, 0} v^{(2)}_{-1, 0} v_{\vphantom{1}}^{(3)}}{p \delta \delta_{-1, 0}  } \nonumber\\
+ \frac{\left(v^{(1)}_{2, 0} v_{\vphantom{1}}^{(2)} v^{(3)}_{1, 0}  +v^{(1)}_{1, 0} v^{(2)}_{1, 0}  v^{(3)}_{2, 0}-3 v^{(2)}_{2, 0} v^{(1)}_{1, 0} v_{\vphantom{1}}^{(3)}  +v_{\vphantom{1}}^{(1)} v^{(2)}_{1, 0}   v^{(3)}_{1, 0}\right) v_{\vphantom{1}}^{(1)} v^{(2)}_{1, 0} v^{(3)}_{-1, 0}  }{p \delta \delta_{-1, 0}  } \nonumber\\
\left.+ \frac{a_1}{p v_{\vphantom{1}}^{(1)}  v^{(3)}_{1, 0} \delta}\right] \nonumber\\
=(D_n - 1) \left[ \frac{q \left(v_{\vphantom{1}}^{(1)} v^{(3)}_{-1, 0}   b_3 -  v^{(1)}_{-1, 0} v_{\vphantom{1}}^{(2)} v_{\vphantom{1}}^{(3)}    b_1 - v^{(2)}_{-1, 0} v_{\vphantom{1}}^{(3)}    b_2\right)}{(p^4-q^4)v^{(1)}_{0, 1} v^{(2)}_{0, 1} v^{(3)}_{1, 0} v^{(3)}_{0, 1}  \delta_{-1, 0}} 
 \right. \nonumber\\
\left. + \frac{3 p^3 v^{(1)}_{0, 1} v^{(2)}_{0, 1} v^{(3)}_{1, 0}  -p^2 q v^{(1)}_{0, 1}v^{(2)}_{1, 0} v^{(3)}_{0, 1}    -  p q^2 v^{(1)}_{1, 0} v^{(2)}_{0, 1} v^{(3)}_{0, 1}  -q^3 v^{(1)}_{0, 1} v^{(2)}_{0, 1} v^{(3)}_{0, 1} }{(p^4-q^4) v^{(1)}_{0, 1} v^{(2)}_{0, 1} v^{(3)}_{1, 0} }\right],\label{ALN3CL2}
\end{eqnarray}
where
\begin{eqnarray*}
\fl a_1 = 2 v_{\vphantom{1}}^{(1)} v^{(1)}_{2, 0} v_{\vphantom{1}}^{(2)} \left(v^{(3)}_{1, 0}\right)^2   - \left(v^{(1)}_{1, 0}\right)^2 v^{(2)}_{1, 0} v_{\vphantom{1}}^{(3)}  v^{(3)}_{2, 0}\\
- v_{\vphantom{1}}^{(1)} v^{(1)}_{1, 0} v^{(2)}_{1, 0}  v_{\vphantom{1}}^{(3)}   v^{(3)}_{1, 0}
+3 v_{\vphantom{1}}^{(1)} v^{(1)}_{1, 0}v^{(2)}_{2, 0}  v_{\vphantom{1}}^{(3)} v^{(3)}_{1, 0}    +\left(v_{\vphantom{1}}^{(1)} \right)^2 v^{(2)}_{1, 0}\left(v^{(3)}_{1, 0}\right)^2 ,\\
\fl b_1 = v^{(1)}_{0, 1} \left(v^{(2)}_{0, 1} \left(v^{(3)}_{1, 0}\right)^2  - v^{(2)}_{1, 0} \left(v^{(3)}_{0, 1}\right)^2\right) p^2\\
+ v^{(3)}_{0, 1} \left( v^{(1)}_{0, 1} v^{(2)}_{1, 0} v^{(3)}_{1, 0} -v^{(1)}_{1, 0}v^{(2)}_{0, 1}  v^{(3)}_{0, 1}\right)  p q
-v^{(2)}_{0, 1} v^{(3)}_{0, 1} \left(v^{(1)}_{0, 1} v^{(3)}_{0, 1}-v^{(1)}_{1, 0} v^{(3)}_{1, 0}\right) q^2,\\
\fl b_2 = v^{(1)}_{0, 1} v^{(3)}_{0, 1} \left(v^{(1)}_{0, 1} v^{(2)}_{0, 1} v^{(3)}_{1, 0} - v^{(1)}_{1, 0} v^{(2)}_{1, 0}v^{(3)}_{0, 1} \right) p^2\\
+v^{(2)}_{0, 1} \left(v^{(1)}_{0, 1} v^{(3)}_{1, 0}-v^{(1)}_{1, 0} v^{(3)}_{0, 1} \right) \left(v^{(1)}_{0, 1} v^{(3)}_{1, 0}+v^{(1)}_{1, 0} v^{(3)}_{0, 1} \right)  pq\\
+v^{(1)}_{0, 1} v^{(3)}_{0, 1} \left(v^{(1)}_{0, 1} v^{(2)}_{1, 0}  v^{(3)}_{1, 0} -v^{(1)}_{1, 0} v^{(2)}_{0, 1}  v^{(3)}_{0, 1}\right) q^2,\\
\fl b_3 =v^{(3)}_{0, 1} \left(v^{(1)}_{0, 1} \left(v^{(2)}_{1, 0}\right)^2  v^{(3)}_{0, 1}-v^{(1)}_{1, 0}  \left(v^{(2)}_{0, 1}\right)^2v^{(3)}_{1, 0}\right) p^2\\
 -v^{(2)}_{0, 1} v^{(3)}_{0, 1} \left(v^{(1)}_{0, 1} v^{(2)}_{0, 1} v^{(3)}_{1, 0}  -v^{(1)}_{1, 0} v^{(2)}_{1, 0} v^{(3)}_{0, 1} \right)  p q\\
 -v^{(1)}_{0, 1} v^{(2)}_{0, 1} \left( v^{(2)}_{0, 1} \left(v^{(3)}_{1, 0}\right)^2-v^{(2)}_{1, 0} \left(v^{(3)}_{0, 1}\right)^2 \right) q^2.
\end{eqnarray*}
And
\begin{eqnarray}
\fl (D_m - 1)\left[\frac{v_{\vphantom{1}}^{(1)}  v^{(2)}_{1, 0} v^{(3)}_{1, 0} \left(-3 v^{(2)}_{2, 0} v^{(1)}_{2, 0} v^{(3)}_{3, 0}+v^{(2)}_{3, 0} v^{(3)}_{1, 0} v^{(1)}_{2, 0}+v^{(1)}_{3, 0} v^{(3)}_{2, 0} v^{(2)}_{1, 0}+v^{(2)}_{2, 0} v^{(1)}_{1, 0} v^{(3)}_{2, 0}\right)}{p \delta_{1, 0} \delta} \right.\nonumber \\
 - \frac{2 v^{(2)}  v^{(1)}_{2, 0} \left(v^{(2)}_{2, 0} v^{(1)}_{2, 0} v^{(3)}_{3, 0}-v^{(2)}_{3, 0} v^{(3)}_{1, 0} v^{(1)}_{2, 0}+v^{(2)}_{2, 0} v^{(1)}_{1, 0} v^{(3)}_{2, 0}-v^{(1)}_{3, 0} v^{(3)}_{2, 0} v^{(2)}_{1, 0}\right) v^{(3)}_{1, 0}}{p \delta_{1, 0} \delta}  \nonumber\\
 - \frac{v^{(3)}  v^{(1)}_{1, 0} \left(v^{(1)}_{3, 0} v^{(3)}_{2, 0} v^{(2)}_{1, 0}+v^{(2)}_{2, 0} v^{(1)}_{2, 0} v^{(3)}_{3, 0}-3 v^{(2)}_{3, 0} v^{(3)}_{1, 0} v^{(1)}_{2, 0}+v^{(2)}_{2, 0} v^{(1)}_{1, 0} v^{(3)}_{2, 0}\right) v^{(2)}_{2, 0}}{p \delta_{1, 0} \delta} \nonumber\\
\left.+ \frac{1}{2} \frac{a_2}{p v^{(1)}  v^{(3)}_{1, 0} \delta}\right] = (D_n -1) \frac{4 v^{(2)}_{1, 0} v^{(1)}_{1, 0} v^{(3)}_{2, 0} v^{(3)}_{0, 1}}{\delta \left(p v^{(3)}_{0, 1}-q v^{(3)}_{1, 0}\right)}, \label{ALN3CL3}
\end{eqnarray}
where
\begin{eqnarray*}
\fl a_2 = -v_{\vphantom{1}}^{(1)}  v^{(1)}_{2, 0} v_{\vphantom{1}}^{(2)} \left(v^{(3)}_{1, 0}\right)^2   -3 v_{\vphantom{1}}^{(1)} v^{(1)}_{1, 0} v^{(2)}_{2, 0} v_{\vphantom{1}}^{(3)}  v^{(3)}_{1, 0}    
-v^{(1)}_{1, 0} v^{(1)}_{2, 0} v_{\vphantom{1}}^{(2)} v_{\vphantom{1}}^{(3)}   v^{(3)}_{1, 0}  \\
 + \left(v^{(1)}_{1, 0}\right)^2 v^{(2)}_{1, 0} v_{\vphantom{1}}^{(3)}  v^{(3)}_{2, 0}
+ v_{\vphantom{1}}^{(1)} v^{(1)}_{1, 0} v^{(2)}_{1, 0}  v_{\vphantom{1}}^{(3)}    v^{(3)}_{1, 0}\\
+\left(v_{\vphantom{1}}^{(1)} \right)^2 v^{(2)}_{1, 0} \left(v^{(3)}_{1, 0}\right)^2 
-\left(v^{(1)}_{1, 0}\right)^2 v^{(2)}_{2, 0} \left(v_{\vphantom{1}}^{(3)} \right)^2 +3 v_{\vphantom{1}}^{(1)} v^{(1)}_{1, 0} v^{(2)}_{1, 0}   v^{(3)}_{1, 0}  v^{(3)}_{2, 0}
\end{eqnarray*}

Recall that $\delta =  v^{(1)}_{1, 0} v^{(2)}_{2, 0} v_{\vphantom{1}}^{(3)}  +v^{(1)}_{2, 0} v_{\vphantom{1}}^{(2)} v^{(3)}_{1, 0}  +v^{(1)}_{1, 0} v^{(2)}_{1, 0}  v^{(3)}_{2, 0}+ v_{\vphantom{1}}^{(1)} v^{(2)}_{1, 0}  v^{(3)}_{1, 0}  $.

Since the equation (\ref{ALN3main}) is invariant under the replacing $n \leftrightarrow m$ one can easily find from the explicit formulas (\ref{ALN3logCL})-- (\ref{ALN3CL3}) conservation laws on the other direction.

Let us give also expression of the first conservation law in terms of the matrix entries of the potentials $f$ and $g$:
\begin{equation}  \label{ALN3logCL2}
(D_m - 1) \log \det_3 P^{(0)} = (D_n - 1) \log \frac{1}{f_{44}- g_{44}}.
\end{equation}
where $\det_3 P^{(0)}$ is the third order leading principal minor of the matrix $P^{(0)}$ defined in (\ref{ALN3P}).

Equations (\ref{ALN1logCL2}), (\ref{ALN2logCL2}) and (\ref{ALN3logCL2}) above show that the logarithmic conservation law is given by formulas which are almost the same for all $N' \geq 1$. It is easily proved that for arbitrary natural $N'$ the relation
\begin{equation*}
(D_m - 1) \log \det_{(N'-1)} P^{(0)} = (D_n - 1) \log \frac{1}{f_{N'N'}- g_{N'N'}},
\end{equation*}
defines a conservation law for the system (\ref{discreteKP}),
where $\det_{(N'-1)} P^{(0)}$ is the $(N'-1)$-th order leading principal minor of the corresponding matrix $P^{(0)}$.

\section{Diagonalization of the Lax pair and construction of the conservation laws for the discrete Tzitzeica equation}

Consider a discrete version of the famous Tzitzeica equation 
\begin{equation} \label{Tz:eq1}
h h_{1,1} \left( c^{-1} h_{1,0} h_{0,1} - h_{1,0} - h_{0,1} \right) + h_{1,1} + h - c = 0.
\end{equation}
found in \cite{Adler,Bobenko-Schief}. Equation (\ref{Tz:eq1}) is the consistency condition of the following system of linear equations 
\begin{equation} \label{Tz:eq2}
y_{1,0} = f y, \qquad y_{0,1} = g y,
\end{equation}
where
\begin{eqnarray*}
f = \left( \begin{array}{ccc}
0 & 1 & 0\\
\frac{c(- h_{1,0}+\lambda^2)}{h_{1,0}(h-c)} & c \frac{h h_{1,0} - 1}{(h-c) h_{1,0}} & \frac{c - \lambda^2}{h-c}h\\
-1 & h & h
\end{array} \right), \\
g = \left( \begin{array}{ccc}
0 & 0 & 1\\
-1 & h & h\\
\frac{c(h_{0,1}c - 1 - \lambda^2(h_{0,1}-c))}{(\lambda^2-c)h_{0,1}(h-c)} & \frac{(1-c^2)h}{(\lambda^2-c)(h-c)} & \frac{c(h h_{0,1}-1)}{h_{0,1}(h-c)}
\end{array} \right).
\end{eqnarray*}

Potential $f$ has two singular points $\lambda=\infty$ and $\lambda=0$. Concentrate on the point $\lambda=\infty$.

Note that the algorithm discussed in the Appendix applied to the first equation in (\ref{Tz:eq2}) leads to the representation (\ref{S3eq12}) for which the factor $P_n(\lambda)$ evaluated due to the formula (\ref{S3eq13}) fails to satisfy the requirement (\ref{S3eq14}). So we cannot immediately apply Theorem 1 to this problem.
In this case we have first to change the leading term for $\lambda\rightarrow\infty$ of the potential $f_n(\lambda)$ by using the so-called cut-off transformation outlined for the differential equations in \cite{Wasow}. The desired transformation consists of changing of the variables as $y = \chi y'$ which results at $y'_{1,0} = f' y'$. Here
\begin{eqnarray*}
\chi = \left( \begin{array}{ccc}
1 & 0 & 0\\
0 & \lambda & 0\\
0 & 0 & 1
\end{array} \right), \\
f':=\chi^{-1} f \chi= \left(\begin{array}{ccc}
0 & \lambda & 0\\
\frac{c(-h_{1,0}+\lambda^2)}{\lambda h_{1,0}(h-c)} & \frac{c(h h_{1,0}-1)}{(h-c)h_{1,0}} & \frac{(c-\lambda^2)h}{\lambda(h-c)}\\
-1 & h \lambda & h
\end{array} \right).
\end{eqnarray*}
Emphasize that the new potential $f'$ grows as $\lambda$ for $\lambda\rightarrow\infty$ while $f$ grows as $\lambda^2$. This explains why the transformation is called ``cut-off''. We factor the matrix $f'$ into a product of the form $f' = \alpha Z \beta$ with
\begin{equation*}
\alpha = \left( \begin{array}{ccc}
0 & 1 & 0\\
\frac{c(-h_{1,0}+\lambda^2)}{\lambda^2h_{1,0}(h-c)} & \frac{c(h h_{1,0}-1)}{\lambda(h-c)h_{1,0}} & 0\\
-\lambda^{-1} & h & h
\end{array} \right), \qquad 
Z = \left( \begin{array}{ccc}
\lambda & 0 & 0\\
0 & \lambda & 0\\
0 & 0 & 1
\end{array} \right),
\end{equation*}
\begin{equation*}
\beta = \left( \begin{array}{ccc}
1 & 0 & \frac{(c-\lambda^2) h h_{1,0}}{c (\lambda^2 - h_{1,0})}\\
0 & 1 & 0\\
0 & 0 & -\frac{(h_{1,0}-c)\lambda^2}{c(\lambda^2-h_{1,0})}
\end{array} \right).
\end{equation*}
Substituting $y' = \beta \psi$, we write the equation in the required form $\psi_{1,0} = P(\lambda)Z\psi$, where $P(\lambda) = \beta_{1,0}\alpha$ or, in
the coordinate representation,
\begin{eqnarray*}
P(\lambda) = P^{(0)} + P^{(1)} \lambda^{-1} + P^{(2)} \lambda^{-2}+ \cdots,\\
P^{(0)} = \left( \begin{array}{ccc}
0 & -\frac{h h_{1,0} h_{2,0}-c}{c} & -\frac{h h_{1,0} h_{2,0}}{c}\\
\frac{c}{(h-c)h_{1,0}} & 0 & 0\\
0 & -\frac{(h_{2,0}-c)h}{c} & -\frac{(h_{2,0}-c)h}{c}
\end{array} \right), \\
P^{(1)} = \left( \begin{array}{ccc}
\frac{h_{1,0} h_{2,0}}{c} & 0 & 0\\
0 & \frac{c(h h_{1,0}-1)}{(h-c)h_{1,0}} & 0\\
\frac{h_{2,0}-c}{c} & 0 & 0
\end{array} \right), \\
P^{(2)} = -\left( \begin{array}{ccc}
0 & \frac{h h_{1,0} h_{2,0}(h_{2,0}-c)}{c} & \frac{h h_{1,0} h_{2,0}(h_{2,0}-c)}{c} \\
\frac{c}{h-c} & 0 & 0\\
0 & \frac{(h_{2,0}-c)h h_{2,0}}{c} &  \frac{(h_{2,0}-c)h h_{2,0}}{c}
\end{array} \right).
\end{eqnarray*}
Function $P(\lambda)$ satisfies the settings of the Theorem 1, indeed  the functions $\det_2P(\lambda=\infty) = \frac{h h_{1,0}h_{2,0}-c}{(h-c)h_{1,0}}$ and $\det P(\lambda=\infty)=\frac{(h_{2,0}-c)h}{(h-c)h_{1,0}}$ do not vanish if the variable $h=h(n,m)$ satisfies the inequalities   $h h_{1,0} h_{2,0} - c \neq 0$, $h-c \neq 0$, $h_{2,0}-c \neq 0$ and $h\neq 0$ for all $n$ and $m$.  Therefore there exist formal series
\begin{eqnarray*}
T = T^{(0)} + T^{(1)} \lambda^{-1} + T^{(2)} \lambda^{-2} + \cdots, \\
H = H^{(0)} + H^{(1)} \lambda^{-1} + H^{(2)} \lambda^{-2} + \cdots,
\end{eqnarray*}
block-diagonalizing the equation $\psi_{1,0} = P(\lambda)Z\psi$ around $\lambda = \infty$ as $\varphi_{1,0} = HZ\varphi$, where $\psi=T \varphi$. 
The coefficients $T^{(k)}$, $H^{(k)}$ are found by solving consecutively the set of equations (\ref{eq24}), (\ref{eq25}), (\ref{eq26}). Let us write the first terms of the series $T$  explicitly:
\begin{eqnarray*}
T^{(0)} = \left( \begin{array}{ccc}
1 & 0 & 0\\
0 & 1 & 0\\
\frac{(h_{1,0}-c)h_{-1,0}}{h_{-1,0} h h_{1,0}-c} & 0 & 1
\end{array} \right), \\
T^{(1)} = \left( \begin{array}{ccc}
0 & 0 & 0\\
0 & 0 & -\frac{h h_{1,0} h_{2,0}}{h h_{1,0} h_{2,0}-c}\\
0 & -\frac{h(h_{1,0}-c)(h_{-1,0}-c)(h_{-1,0}h_{-2,0}-1)}{(h_{-1,0} h h_{1,0}-c)(h_{-2,0} h_{-1,0} h-c)} & 0
\end{array} \right),\\
T^{(2)} = \left( \begin{array}{ccc}
0 & 0 & \frac{h h_{1,0} h_{2,0}\left( c^2 w - c h_{1,0} h_{3,0} v + h h_{1,0} h_{3,0} u_{2,0}\right)}{ v v_{1,0}}\\
0 & 0 & 0\\
\frac{cu_{1,0} \rho}{v^2_{-1,0} v_{-2,0} v_{-3,0}} & 0 & 0
\end{array} \right),
\end{eqnarray*}
where 
\begin{eqnarray*}
\fl \rho =  -h_{-1,0}h_{1,0}v_{-3,0}v_{-2,0} + h_{-2,0} h^2_{-1,0}h v_{-3,0}v_{-1,0} - c h^2_{-1,0}h h_{1,0} v_{-3,0} + h_{-3,0}h_{-2,0}h_{-1,0}v_{-2,0}  \\+ c^2 h^2_{-1,0}w_{-3,0}-c^2 w_{-2,0}+ c h h_{-1,0} u_{-1,0} - h_{-2,0} h^2_{-1,0} u.
\end{eqnarray*}
Here and hereafter $u = h-c$, $v = h h_{1,0} h_{2,0} -c$ and $w = h h_{1,0} - 1$.

Let us write the first terms of the series $H$  explicitly:
\begin{eqnarray*}
H^{(0)} = \left( \begin{array}{ccc}
0 & \frac{c - h h_{1,0} h_{2,0}}{c} & 0\\
\frac{c}{(h-c)h_{1,0}} & 0 & 0\\
0 & 0 & \frac{(h_{2,0}-c)h}{h h_{1,0} h_{2,0}-c}
\end{array} \right),\\
H^{(1)} = \left( \begin{array}{ccc}
\frac{h_{1,0}h_{2,0}(h h_{-1,0}-1)}{h_{-1,0} h h_{1,0}-c} & 0 & 0\\
0 & \frac{c(h h_{1,0}-1)}{(h-c)h_{1,0}} & 0\\
0 & 0 & 0
\end{array} \right),
\end{eqnarray*}
where 
\begin{equation*}
\nu =   c^2(h h_{1,0}-1) - c h_{1,0} h_{3,0} (h h_{1,0} h_{2,0} - c) + h h_{1,0} h_{3,0} (h_{2,0} - c).
\end{equation*}  
\begin{equation*}
H^{(2)} = \left( \begin{array}{ccc}
0 & \frac{h h_{1,0} h_{2,0} a}{c v_{-1,0} v_{-2,0}} & 0\\
-\frac{c}{u} & 0 & 0\\
0 & 0 & \frac{h h_{2,0} u_{2,0} b}{ v^2 v_{1,0} v_{2,0}}
\end{array} \right),
\end{equation*}
where 
\begin{eqnarray*}
\fl a = -h_{2,0} v_{-1,0} v_{-2,0} - c^2 v_{-1,0} - h u_{-1,0} u_{1,0} \\
+ c h h^2_{-1,0} h_{-2,0} w + h h_{1,0} h_{-1,0} h_{-2,0} u_{-1,0},\\
\fl b =  h_{1,0}h_{3,0} (h_{2,0} - h_{1,0}) v v_{2,0} - h h_{1,0} h_{3,0} u_{2,0}u_{1,0} 
 + h_{2,0} h_{3,0} h_{4,0} v u_{1,0} - c^2 u_{1,0} w.
\end{eqnarray*}

The operator $M = D^{-1}_m g$ is diagonalized (in fact it is ``block diagonalized'') as $M_0 = D^{-1}_m S$, where
\begin{equation*}
S = D_m \left( T^{-1} \beta \chi^{-1} \right)M \chi \beta^{-1} T = S^{(0)} + S^{(1)} \lambda^{-1} + S^{(2)} \lambda^{-2} + \cdots.
\end{equation*}
We have
\begin{equation*}
S^{(0)} = \left( \begin{array}{ccc}
\frac{c(h_{0,1}-c)}{c(h h_{1,0}-1) - h h_{0,1}(h_{1,0}-1)} & 0 & 0\\
0 & h & 0\\
0 & 0 & \frac{c h \left(h_{0,1} (h_{1,0}-c)+1 - c h_{1,0}\right)}{(h_{1,0}-c)(h-c)h_{0,1}}
\end{array} \right),
\end{equation*}
\begin{equation*}
S^{(1)} = \left( \begin{array}{ccc}
0 & \frac{h h_{0,1} (c^2 - 1)}{c(h h_{1,0} - 1) - h h_{0,1} (h_{1,0} - c)} & 0\\
-\frac{(h_{-1,0}h-1)c}{h_{-1,0} h h_{1,0}-c} & 0 & 0\\
0 & 0 & 0
\end{array} \right),
\end{equation*}
\begin{equation*}
S^{(2)} = \left( \begin{array}{ccc}
-\frac{c A}{v_{-1,0}(cw-h h_{0,1}u_{1,0})} & 0 & 0\\
0 & -\frac{u_{1,0} u_{-1,0} w_{-2,0} h^2}{v_{-1,0} v_{-2,0}} & 0\\
0 & 0 & -\frac{h c B (h_{1,0} c - 1 - h_{0,1}u)}{ u v v_{1,0} h_{0,1} (cw - h h_{0,1} u_{1,0})}
\end{array} \right),
\end{equation*}
where
\begin{eqnarray*}
\fl A = -c^3 v_{-1,0} - c^2 h h_{0,1} w_{-1,0} + c h h_{0,1} h_{-1,0} h^2_{1,0} w \\
+ c^2 h^2 h^2_{1,0} h_{-1,0} u_{0,1} u_{1,0} + c^2 h u_{-1,0} (h_{1,0} h_{0,1} + 1) \\
+u_{1,0} \left( - c^2 h_{-1,0} h h_{0,1} w - c^2 h h^2_{0,1}  w_{-1,0} - h^2 h_{-1,0} h_{1,0}h^2_{0,1} u_{1,0}  \right.\\
+ c^2 h h_{-1,0} h_{1,0} - c^2 h^2 h_{1,0} h_{-1,0} h_{0,1} - c h h_{0,1} \left. \right),\\
\fl B = c v v_{1,0} + (h_{1,0}h_{3,0} - 1)u u_{0,1} h h_{1,0} h_{2,0} +h^2 h_{0,1} h_{1,0} h_{2,0} u_{1,0} v_{1,0} + c h v_{1,0} (1 - h h^2_{1,0} h_{2,0}).
\end{eqnarray*}
Thus we have an infinite sequence of the conservation laws for equation (\ref{Tz:eq1}) obtained from (see (\ref{S3eq7}))
\begin{equation*}
(D_n - 1) \log \det S_{ii} = (D_m - 1) \log \det H_{ii}.
\end{equation*}
Let us give two of them:
\begin{eqnarray}
\fl (D_m - 1) \log \frac{(h_{2,0} - c)h}{h h_{1,0} h_{2,0} - c}  \nonumber\\
=(D_n - 1) \log \frac{c h \left(h_{0,1} (h_{1,0}-c)+1 - c h_{1,0}\right)}{(h_{1,0}-c)(h-c)h_{0,1}}, \label{Tz:eq3}\\
 (D_m - 1) \frac{ h_{2,0} b}{ v v_{1,0} v_{2,0} } 
= (D_n -1) \frac{B u_{1,0} }{v v_{1,0} (h h_{0,1}u_{1,0} - c w)}. \label{Tz:eq4}
\end{eqnarray}
Recall that $u = h-c$, $v = h h_{1,0} h_{2,0} -c$, $w = h h_{1,0} - 1$,
\begin{eqnarray*}
\fl b =  h_{1,0}h_{3,0} (h_{2,0} - h_{1,0}) v v_{2,0} - h h_{1,0} h_{3,0} u_{2,0}u_{1,0} 
 + h_{2,0} h_{3,0} h_{4,0} v u_{1,0} - c^2 u_{1,0} w,\\
\fl B = c v v_{1,0} + (h_{1,0}h_{3,0} - 1)u u_{0,1} h h_{1,0} h_{2,0} +h^2 h_{0,1} h_{1,0} h_{2,0} u_{1,0} v_{1,0} + c h v_{1,0} (1 - h h^2_{1,0} h_{2,0}).
\end{eqnarray*}

Since the equation (\ref{Tz:eq1}) is invariant under the change of independent variables $n \leftrightarrow m$ then by applying this change to the sequence of the conservation laws found we obtain one more sequence of the conservation laws. Exact formulas (\ref{Tz:eq3}), (\ref{Tz:eq4}) give rise to:
\begin{eqnarray}
\fl (D_n - 1) \log \frac{(h_{0,2} - c)h}{h h_{0,1} h_{0,2} - c}  \nonumber\\
=(D_m - 1) \log \frac{c h \left(h_{1,0} (h_{0,1}-c)+1 - c h_{0,1}\right)}{(h_{0,1}-c)(h-c)h_{1,0}}, \label{Tz:eq5} \\
(D_n - 1) \frac{ h_{0,2} \bar{b}}{ \bar{v} \bar{v}_{0,1} \bar{v}_{0,2} } 
= (D_m -1) \frac{\bar{B} u_{0,1} }{\bar{v} \bar{v}_{0,1} (h h_{1,0}\bar{u}_{0,1} - c \bar{w})}, \label{Tz:eq6}
\end{eqnarray}
where  $\bar{u} = h-c$, $\bar{v} = h h_{0,1} h_{0,2} -c$, $\bar{w} = h h_{0,1} - 1$,
\begin{eqnarray*}
\fl \bar{b} =  h_{0,1}h_{0,3} (h_{0,2} - h_{0,1}) \bar{v} \bar{v}_{0,2} - h h_{0,1} h_{0,3} \bar{u}_{0,2}\bar{u}_{0,1} 
 + h_{0,2} h_{0,3} h_{0,4} \bar{v} \bar{u}_{0,1} - c^2 \bar{u}_{0,1} \bar{w},\\
\fl \bar{B} = c \bar{v} \bar{v}_{0,1} + (h_{0,1}h_{0,3} - 1)\bar{u} \bar{u}_{1,0} h h_{0,1} h_{0,2} +h^2 h_{1,0} h_{0,1} h_{0,2} \bar{u}_{0,1} \bar{v}_{0,1} + c h \bar{v}_{0,1} (1 - h h^2_{0,1} h_{0,2}).
\end{eqnarray*}

The conservation laws (\ref{Tz:eq3}), (\ref{Tz:eq5}) have been found earlier (see eq. (12) in \cite{Adler}). It is shown in \cite{mx14} that equation (\ref{Tz:eq1}) admits a further reduction to a nonlinear integrable quad equation and these two conservation laws (\ref{Tz:eq3}), (\ref{Tz:eq5}) convert to the conservation laws of the reduced equation.

\section{A multi-quadratic integrable quad equation}

In this section we consider the following quad equation
\begin{eqnarray}	\label{S6eq2dashed}
\fl \sqrt{(u_{1,1}-u_{0,1})u_{0,1}} = \varepsilon^{-1} \sqrt{(u_{1,0}-u)u} 
-\sqrt{(1+\varepsilon^{-2})u u_{0,1} - \varepsilon^{-2} u^2 - u^2_{0,1}},
\end{eqnarray}
where $\varepsilon$ is a parameter. Equation (\ref{S6eq2dashed}) is the consistency condition of the following overdetermined system of equations
\begin{equation}	\label{S6eq10}
y_{1,0} = f y, \qquad y_{0,1} = gy
\end{equation}
where
\begin{eqnarray}
f = \left( \begin{array}{cc}
\lambda \sqrt{\frac{u}{u_{1,0}}}  & \sqrt{1 - \frac{u}{u_{1,0}}}\\
-\lambda \sqrt{1 - \frac{u}{u_{1,0}}}  & \sqrt{\frac{u}{u_{1,0}}} 
\end{array} \right),	\label{S6eq8}\\
g = \left( \begin{array}{cc}
\lambda \sqrt{\frac{u}{u_{0,1}}} + \varepsilon \sqrt{\frac{u_{0,1}}{u}} & \varepsilon \mu\\
-\varepsilon \lambda \mu & \varepsilon \lambda \sqrt{\frac{u_{0,1}}{u}} + \sqrt{\frac{u}{u_{0,1}}}
\end{array} \right),				\label{S6eq9}
\end{eqnarray}
where $\mu = \sqrt{1 + \varepsilon^{-2} - \varepsilon^{-2} \frac{u}{u_{0,1}} - \frac{u_{0,1}}{u}}$. 

Note that equation (\ref{S6eq2dashed}) can easily be rewritten in a polynomial form 
\begin{eqnarray}
\fl 2 \varepsilon^2 u u_{0,1}(
u_{1,0}  u_{1,1} +  u u_{1,0} + u_{0,1} u_{1,1} + u u_{0,1})-4 \varepsilon^2 u u_{0,1} (u_{1,0} u_{0,1} + u u_{1,1})-\nonumber\\ 
 -  \varepsilon^4  u^2_{0,1}(u_{1,1}-u)^2 + 2 u^2 u_{1,0} u_{0,1} - u^2 (u^2_{1,0} + u^2_{0,1}) = 0.	\label{S6eq2}
\end{eqnarray}
Here the left hand side is a quadratic polynomial with respect to each of its arguments.
Such kind equations are called multi-quadratic.  

We describe the conservation laws for equation (\ref{S6eq2dashed}) by applying the formal diagonalization method. Begin with the first equation in the Lax pair. Potential $f$ has two singular points $\lambda=0$ and $\lambda=\infty$. Due to the involution $f^{-1*}(\lambda)=f(\lambda^{-1*})$ of the potential the eigenfunction has similar asymptotic behavior around these two singularities. We concentrate on one of them, taking $\lambda_0=\infty$.
Obviously potential $f$ is of the necessary special form (\ref{MainSpecialForm}). 
Indeed we have $f = P Z$, where 
\begin{equation*}
P = \left( \begin{array}{cc}
\sqrt{\frac{u}{u_{1,0}}} & \sqrt{1 - \frac{u}{u_{1,0}}}\\
-\sqrt{1 - \frac{u}{u_{1,0}}} & \sqrt{\frac{u}{u_{1,0}}} 
\end{array} \right), \qquad
Z = \left( \begin{array}{cc}
\lambda & 0\\
0 & 1
\end{array} \right).
\end{equation*}
The leading principal minor $\det_1P=\sqrt{\frac{u}{u_{1,0}}} $ and the determinant $\det P=1$ of the matrix $P$ do not vanish if $u_{n,m}\neq0$. Therefore all the conditions of Theorem 1 hold. Omitting the tedious computations we give only the answers. The coefficients of the formal series 
\begin{equation*}
T = T^{(0)} + T^{(1)} \lambda^{-1}+ \cdots, \qquad h = h^{(0)} + h^{(1)} \lambda^{-1} + \cdots
\end{equation*}
are 
\begin{eqnarray*}
h^{(0)} = \left( \begin{array}{cc}
\sqrt{\frac{u}{u_{1,0}}} & 0\\
0 & \sqrt{\frac{u_{1,0}}{u}}
\end{array} \right), \qquad
T^{(0)} = \left( \begin{array}{cc}
1 & 0 \\
- \sqrt{\frac{v_{-1,0}}{u_{-1,0}}} & 1
\end{array} \right)
\end{eqnarray*}
where $v := u_{1,0} - u$ and
\begin{eqnarray*}
h^{(1)} = \left(\begin{array}{cc}
-\sqrt{\frac{v_{-1,0}v}{u_{-1,0}u}} & 0\\
0 & \frac{\sqrt{v_{1,0} v}}{u}
\end{array}  \right), \qquad
T^{(1)} = \left( \begin{array}{cc}
0 & -\sqrt{\frac{v}{u}}\\
- \sqrt{\frac{v_{-2,0}}{u_{-2,0}}} \frac{u}{u_{-1,0}} & 0
\end{array} \right), 
 \end{eqnarray*}

\begin{eqnarray*}
h^{(2)} = \left( \begin{array}{cc}
-\sqrt{\frac{v v_{-2,0}}{u_{1,0} u_{-2,0}}} \frac{u}{u_{-1,0}} & 0\\
0 & \frac{\sqrt{v u_{1,0}}}{u} \left( \sqrt{\frac{v}{u}} \frac{v_{1,0}}{u_{1,0}} + \sqrt{\frac{v_{2,0}}{u_{2,0}}} \frac{u_{2,0}}{u_{1,0}} \right)
\end{array} \right),\\
T^{(2)} = \left( \begin{array}{cc}
0 & -\frac{\sqrt{ v_{1,0} u_{1,0}}}{u}\\
-\frac{u}{u_{-1,0}} \left( \sqrt{\frac{v_{-3,0}}{u_{-3,0}}} \frac{u_{-1,0}}{u_{-2,0}} + \sqrt{\frac{v_{-1,0}}{u_{-1,0}}} \frac{v_{-2,0}}{u_{-2,0}} \right)
\end{array}  \right).
\end{eqnarray*}

Recall that the change of the variables $y=T\varphi$ reduces the first equation in (\ref{S6eq10}) to the diagonal form $\varphi_{1,0}=hZ\varphi$ and the second one to the form $\varphi_{0,1}=S\varphi$ where $S$ is a formal series
\begin{equation*}
S = S^{(0)} \lambda + S^{(1)} + S^{(2)} \lambda^{-1} + \cdots.
\end{equation*}
We found several  members of the series explicitly
\begin{eqnarray*}
S^{(0)} =  \left( \begin{array}{cc}
\sqrt{\frac{u}{u_{0,1}}} & 0\\
0 & \varepsilon \sqrt{\frac{u_{0,1}}{u}}
\end{array} \right),\\
S^{(1)} = \left( \begin{array}{cc}
\varepsilon \sqrt{\frac{u_{0,1}}{u}} \left( \sqrt{\frac{v_{1,0} v_{-1,0}}{u_{1,0} u_{-1,0}}} +1 \right)-\sqrt{\frac{v v_{-1,0}}{u_{-1,0} u_{0,1}}} & 0\\
0 & S^{(1)}_{22}
\end{array} \right),\\
S^{(2)} = \sqrt{\frac{u}{u_{0,1}}} \left( \begin{array}{cc}
S^{(2)}_{11} & 0\\
0 & S^{(2)}_{22}
\end{array} \right),
\end{eqnarray*}
where
\begin{eqnarray*}
S^{(1)}_{22} = \frac{\varepsilon \sqrt{\frac{u_{0,1}}{u}}}{\sqrt{\frac{v_{0,1}}{u_{0,1}}} - \sqrt{\frac{v_{1,0}}{u_{1,0}}}} \left( \sqrt{\frac{v}{u}} \left( 1+ \sqrt{\frac{v_{1,0}v_{0,1}}{u_{1,0}u_{0,1}}}  \right) - \sqrt{\frac{v_{1,1}}{u_{1,1}}} \frac{u_{1,1}}{u_{0,1}}  \right),\\
\fl S^{(2)}_{11} = -\frac{\sqrt{\frac{v_{-2,0}}{u_{-2,0}}} \frac{u}{u_{-1,0}}}{\frac{u_{2,0}}{u_{1,0}} \left( \sqrt{\frac{v}{u}\vphantom{\frac{v_{1,1}}{u_{1,1}}}} - \sqrt{\frac{v_{1,1}}{u_{1,1}}}  \right)} \\
\times \left(  \frac{u_{1,0}}{u} - \frac{u_{2,0}}{u_{1,0}} + \sqrt{\frac{v_{1,0} v_{0,1}}{u_{1,0} u_{0,1}}} \frac{u_{1,0}}{u} - \sqrt{\frac{v v_{1,1}}{u u_{1,1}}} \frac{u_{2,0}}{u_{1,0}}  \right)  ,\\
\fl S^{(2)}_{22}= \frac{\sqrt{\frac{v_{1,0}}{u_{1,0}}} \frac{u_{1,0}}{u}}{\frac{u_{2,0}}{u_{1,0}} \left(  \sqrt{\frac{v}{u}\vphantom{\frac{v_{1,1}}{u_{1,1}}}} - \sqrt{\frac{v_{1,1}}{u_{1,1}}}\right)}\\
\times\left(  \frac{u_{1,0}}{u} - \frac{u_{2,0}}{u_{1,0}} + \sqrt{\frac{v_{1,0} v_{0,1}}{u_{1,0} u_{0,1}}} \frac{u_{1,0}}{u} - \sqrt{\frac{v v_{1,1}}{u u_{1,1}}} \frac{u_{2,0}}{u_{1,0}}  \right). 
\end{eqnarray*}

Now formula (\ref{S3eq7}) allows one to write down the conservation laws. Give two first of them:
\begin{eqnarray*}
\left(D_n - 1 \right) \frac{ \sqrt{\frac{v}{u}} \left( 1+ \sqrt{\frac{v_{1,0}v_{0,1}}{u_{1,0}u_{0,1}}}  \right) - \sqrt{\frac{v_{1,1}}{u_{1,1}}} \frac{u_{1,1}}{u_{0,1}}   } {\sqrt{\frac{v_{0,1}}{u_{0,1}}} - \sqrt{\frac{v_{1,0}}{u_{1,0}}}}
 = \left( D_m -1 \right)(\sqrt{\frac{v v_{1,0}}{u u_{1,0}}}),\\
\left(D_n - 1 \right) \frac{a}{2\left( \sqrt{\frac{v_{0,1}}{u_{0,1}}} - \sqrt{\frac{v_{1,0}}{u_{1,0}}}  \right)^2} = \left( D_m -1 \right) \left( \sqrt{\frac{v v_{2,0}}{u u_{2,0}}} \frac{u_{2,0}}{u_{1,0}} + \frac{1}{2} \frac{v v_{1,0}}{u u_{1,0}} \right).
\end{eqnarray*}
Here
\begin{eqnarray*}
\fl a = \frac{v}{u} \left( \frac{v_{1,0} v_{0,1}}{u_{1,0} u_{0,1}} - 2 \frac{v_{1,0}}{u_{1,0}} \left(  \sqrt{\frac{v_{1,0} v_{0,1}}{u_{1,0} u_{0,1}}} +1 \right) + 1 \right) \\
+2 \sqrt{\frac{v v_{1,1}}{u u_{1,1}}} \frac{u_{2,0} u_{1,1}}{u_{1,0} u_{0,1}} - 2 \sqrt{\frac{v_{1,0} v_{0,1}}{u_{1,0} u_{0,1}}} \left( \sqrt{\frac{v_{0,1}}{u_{0,1}}} + \sqrt{\frac{v_{1,0}}{u_{1,0}}} \right)^2 - \frac{v_{1,1}}{u_{0,1}}.
\end{eqnarray*}

Now let us find conservation laws in the other direction. Consider the second equation in (\ref{S6eq10}). Introduce a new spectral parameter by setting $\eta = \frac{\lambda \varepsilon + 1}{\lambda + \varepsilon}$ then $\lambda = \frac{1- \varepsilon \eta}{\eta - \varepsilon}$. Changing the variable as $y = (\eta - \varepsilon)^{-m} \varphi$, we obtain the equation $\varphi_{0,1} = G \varphi$, where
\begin{equation*}
G(\eta) = \left( \begin{array}{cc}
\eta \varepsilon \left( t - \frac{1}{t} \right) + \frac{1}{t} - \varepsilon^2 t & \varepsilon \mu \eta - \varepsilon^2 \mu\\
\eta \varepsilon^2 \mu - \varepsilon \mu & \eta \left( \frac{1}{t} - \varepsilon^2 t \right) + \varepsilon \left(t - \frac{1}{t}\right)
\end{array} \right).
\end{equation*}
Here $t: = \sqrt{u_{0,1}/u}$. We factor the matrix $G$ into a product of the form $G = \alpha Z \beta$
\begin{eqnarray*}
\alpha = \left( \begin{array}{cc}
\varepsilon \left(t - \frac{1}{t}\right) + \eta^{-1} \left( \frac{1}{t} - \varepsilon^2 t\right) & 0\\
\varepsilon \kappa \left(t - \frac{1}{t}\right)\left( \varepsilon - \eta^{-1} \right) & \frac{\eta (1 - \varepsilon^2)^2}{\varepsilon \eta \left(  t - \frac{1}{t} \right) + \left( \frac{1}{t} - \varepsilon^2 t \right)}
\end{array}  \right), \\
Z = \left( \begin{array}{cc}
\eta & 0\\
0 & 1
\end{array} \right), \qquad 
\beta = \left( \begin{array}{cc}
1 & \frac{\kappa(\eta - \varepsilon)}{\eta + \varepsilon \kappa^2}\\
0 & 1
\end{array}  \right),
\end{eqnarray*}
where $\kappa = \mu/(t - t^{-1})$.
Substituting $\varphi = \beta \psi$, we write the equation in the required form $\varphi_{0,1} = P(\eta) Z \varphi$, where 
\begin{equation*}
P(\eta) = \beta_{0,1} \alpha = P^{(0)} + P^{(1)} \eta^{-1} + \cdots.
\end{equation*}
The matrix $P^{(0)} = P(\infty)$ has the form
\begin{equation*}
P^{(0)} = \left( \begin{array}{cc}
\varepsilon \left( t - \frac{1}{t} \right)(1 + \varepsilon \kappa \kappa_{0,1}) & \frac{(1-\varepsilon^2)^2 \kappa_{0,1}}{\varepsilon \left( t -\frac{1}{t} \right)}\\
\varepsilon^2 \kappa \left( t - \frac{1}{t} \right) & \frac{(1 - \varepsilon^2)^2}{\varepsilon \left( t - \frac{1}{t} \right)}
\end{array} \right).
\end{equation*}
and evidently satisfies the conditions of the Theorem 1. Thus the operator $M = D^{-1}_m PZ$ is diagonalizable.
Solving the equation $D_m\left({T\vphantom{ \overline{T}}}^{(0)}\right){h\vphantom{ \overline{T}}}^{(0)} = {P\vphantom{ \overline{T}}}^{(0)} \overline{T}^{(0)}$, we obtain
\begin{eqnarray*}
h^{(0)} = \left( \begin{array}{cc}
\varepsilon \left( t - \frac{1}{t} \right)(1 + \varepsilon \kappa \kappa_{0,1}) & 0\\
0 & \frac{(1-\varepsilon^2)^2}{\varepsilon \left( t - \frac{1}{t} \right)(1 + \varepsilon \kappa \kappa_{0,1})}
\end{array} \right), \\
T^{(0)} = \left( \begin{array}{cc} 
1 & 0\\
\frac{\varepsilon \kappa_{0,-1}}{1 + \varepsilon \kappa \kappa_{0,-1}} & 1
\end{array} \right).
\end{eqnarray*}

The operator $L = D^{-1}_n f$ is reduced to a diagonal form $L_0 = D^{-1}_n S$, where\\ $S:=D_n((\beta^{-1}T)^{-1}) f \beta^{-1} T = S^{(0)} + S^{(1)} \eta^{-1} + \cdots$. The matrix $S^{(0)}$ has the following form:
\begin{eqnarray*}
\fl S^{(0)} =  \sqrt{\frac{u}{u_{1,0}}} \\
\times \left( \begin{array}{cc}
\varepsilon \frac{(\kappa_{1,0} + \kappa_{0,-1})\sqrt{\frac{u_{1,0}}{u}-1}  + \kappa_{1,0}\kappa_{0,-1}  - 1}{1 + \varepsilon \kappa \kappa_{0,-1}} & 0\\
0 &  \frac{-\varepsilon (\kappa + \kappa_{1,-1})\sqrt{\frac{u_{1,0}}{u}-1} + 1  - \varepsilon^2 \kappa \kappa_{1,-1}}{1 + \varepsilon \kappa_{1,0} \kappa_{1,-1}}
\end{array}  \right).
\end{eqnarray*}

We present one conservation law, the others are extremely large
\begin{eqnarray*}
\fl \left(D_n -1  \right) \ln \left( \frac{u_{0,1}}{u} - 1\right) (1 + \varepsilon \kappa \kappa_{0,1}) \\
= \left(D_m - 1  \right) \ln \frac{(\kappa_{1,0} + \kappa_{0,-1})\sqrt{\frac{u_{1,0}}{u}-1}  + \kappa_{1,0}\kappa_{0,-1}  - 1}{1 + \varepsilon \kappa \kappa_{0,-1}}.
\end{eqnarray*}
Recall that $\kappa = \frac{\sqrt{(1+\varepsilon^{-2})u u_{0,1} - \varepsilon^{-2} u^2 - u^2_{0,1}}}{u_{0,1} - u}$.

We construct generalized symmetries of the equation (\ref{S6eq2dashed}):

\begin{equation*}
\frac{{\rm d} u}{{\rm d} \tau_1} = 4 u \sqrt{\left( \frac{u_{1,0}}{u} - 1 \right)\left( \frac{u}{u_{-1,0}} - 1 \right)},
\end{equation*}
and
\begin{equation*}
\frac{{\rm d} u}{{\rm d} \tau_2} = 4(\varepsilon^{-1} - \varepsilon)\frac{u}{1+ \varepsilon  \kappa\kappa_{0,-1}}
\end{equation*}
with $\kappa = \frac{\sqrt{(1+\varepsilon^{-2})u u_{0,1} - \varepsilon^{-2} u^2 - u^2_{0,1}}}{u_{0,1} - u}$   having the Lax pairs
\begin{equation} \label{S6:Lax1}
\frac{d y}{d \tau_1} = A_{(1)} y, \qquad y_{1,0} = f y.
\end{equation}
and 
\begin{equation} \label{S6:Lax2}
y_{0,1} = g y, \qquad \frac{dy}{d \tau_2} = A_{(2)}y.
\end{equation}
where $f$ and $g$ are given by (\ref{S6eq8}) and (\ref{S6eq9}) and 
\begin{eqnarray*}
\fl A_{(1)} = \left( \begin{array}{cc}
\lambda - \lambda^{-1} & 2 \left( \sqrt{\frac{u_{1,0}}{u} - 1} + \sqrt{\frac{u}{u_{-1,0}} - 1} \lambda^{-1} \right)\\
-2 \left( \sqrt{\frac{u}{u_{-1,0}} - 1} \lambda + \sqrt{\frac{u_{1,0}}{u} - 1} \right) & -\lambda + \lambda^{-1}
\end{array} \right),\\
\fl A_{(2)} = \frac{1}{1+\varepsilon \kappa \kappa_{0,-1}} \\
\times
\left(\begin{array}{cc}
(1 - \varepsilon \kappa \kappa_{0,-1})(\eta - \eta^{-1}) & 2 (\kappa \eta - \varepsilon \kappa + \kappa_{0,-1} -  \varepsilon \kappa_{0,-1} \eta^{-1})\\
2 (\varepsilon \kappa_{0,-1} \eta + \varepsilon \kappa - \kappa_{0,-1} -  \kappa \eta^{-1}) & (1 - \varepsilon \kappa \kappa_{0,-1}) (\eta^{-1} - \eta)
\end{array} \right).
\end{eqnarray*}

\section{Appendix. Algorithm of finding  special form of the linear discrete equation}

The problem of finding asymptotic expansion in a parameter around a singularity point for the fundamental solution of the discrete equation (\ref{eq1}) is rather complicated. Even for comparatively well studied systems of linear differential equations the similar problem does not have a complete solution. Theorem 1 shows that the key step in solving the problem in the discrete case is to reduce equation (\ref{eq1}) to the special form (\ref{MainSpecialForm}) at the vicinity of the singular point $\lambda = \lambda_0$ (see \cite{HabLOMI,HabYang}). Below we discuss a method for searching such a special form for a given equation (\ref{eq1}).

Let us fix a block representation of the potential $f = \{ f_{ij} \}_{i,j=1}^{r'}$ such that the block $f_{ij}$ is a matrix of the size $e'_i \times e'_j$, where the positive integers $e'_1, e'_2,\ldots, e'_{r'}$ are chosen as small as possible but satisfying the conditions: $\det f_{ii}$ is not identically zero for $i=1,2,\ldots,r'$ (alternatively, $\det (f^{-1})_{ii}$ is not identically zero). Then find a block lower-triangular $A_L$ (or $B_L$) and a block upper triangular $B_U$ (or $A_U$) such that 
\begin{equation}	\label{S3eq8}
f = A_L B_U \quad(\mbox{ or } f = A_U B_L).
\end{equation}

Suppose now that there exist diagonal matrices $Z_1$ and $Z_2$:
\begin{eqnarray}	
Z_1 = \left( \begin{array}{cccc}
(\lambda - \lambda_0)^{\varepsilon_1} E'_1 & 0 & \ldots & 0\\
0 & (\lambda - \lambda_0)^{\varepsilon_2} E'_2 & \ldots & 0\\
\vdots & \vdots & \ddots & \vdots\\
0 & 0 & \ldots & (\lambda - \lambda_0)^{\varepsilon_{r'}} E'_{r'}
\end{array} \right), \label{S3eq9}\\
Z_2 = \left( \begin{array}{cccc}
(\lambda - \lambda_0)^{\delta_1} E'_1 & 0 & \ldots & 0\\
0 & (\lambda - \lambda_0)^{\delta_2} E'_2 & \ldots & 0\\
\vdots & \vdots & \ddots & \vdots\\
0 & 0 & \ldots & (\lambda - \lambda_0)^{\varepsilon_{r'}} E'_{r'}
\end{array} \right), \label{S3eq10}
\end{eqnarray}
where $\varepsilon_j$, $\delta_j$ for $j = \overline{1, r'}$ are integer and $E'_j$ is the unity matrix of the size $e'_j$, such that the matrix-valued functions $A'(\lambda) = A_L Z^{-1}_1$ and $B'(\lambda) = Z^{-1}_2 B_U$ (respectively $A'(\lambda) = A_U Z^{-1}_1$ and $B'(\lambda) = Z^{-1}_2 B_L$) are analytic and nondegenerate at the vicinity of $\lambda_0$. Evidently $Z' = Z_1 Z_2$ is a diagonal matrix having at the main diagonal powers of $\lambda - \lambda_0$ with integer exponents. By using a linear transformation $Z:=S Z' S^{-1}$ with constant matrix $S$ bring $Z'$ to the form
\begin{equation} 	\label{S3eq11}
Z = \left( \begin{array}{cccc}
(\lambda - \lambda_0)^{\gamma_1} E_1 & 0 & \ldots & 0\\
0 & (\lambda - \lambda_0)^{\gamma_2} E_2 & \ldots & 0\\
\vdots & \vdots & \ddots & \vdots\\
0 & 0 & \ldots & (\lambda - \lambda_0)^{\gamma_{r}} E_{r}
\end{array} \right),
\end{equation}
where the exponents are ordered as follows $\gamma_1 < \gamma_2 < \ldots < \gamma_{r}$ and $E_j$ is the unity matrix of the size $e_j$. Stress that generally speaking the integers $e_1, e_2, \ldots, e_{r}$
define the block representation of a matrix which differs from that defined by the numbers $e'_1, e'_2,\ldots,e'_{r'}$. Formulas (\ref{S3eq8}) give rise to the following representation for the potential $f$
\begin{equation}	\label{S3eq12}
f = \alpha_n(\lambda) Z \beta_n(\lambda)
\end{equation}
which coincides with the necessary representation (\ref{eq11}). Here $\alpha_n(\lambda) = A'(\lambda) S^{-1}$ and $\beta_n(\lambda) = S B'(\lambda)$. Define
\begin{equation}	\label{S3eq13}
P_n(\lambda) = D_n (\beta_n(\lambda)) \alpha_n(\lambda).
\end{equation}
Suppose that the leading principal minors of the matrix $P_n(\lambda_0)$ satisfy the conditions

\begin{equation}	 	\label{S3eq14}
\det_j P_n(\lambda_0) \neq 0 \mbox{ for } j = e_1, e_1 + e_2, e_1+e_2+e_3, \ldots, r.
\end{equation}
Then all conditions of the Theorem 1 are satisfied and therefore equation (\ref{eq1}) is diagonalized
 around $\lambda = \lambda_0$. This case is illustrated with examples of the quad systems  related to the affine Lie algebra $A^{(1)}_N$, $N=1$, $N=2$, $N=3$ (see section 4).

However it might happen that some of the requirements (\ref{S3eq14}) are violated. In this case following the idea of Turritin \cite{Turr} (see also monography \cite{Wasow}) suggested for studying differential equations we apply to the equation (\ref{eq1}) a linear transformation $y=F_n(\lambda)\tilde y$  which is called in \cite{Turr}, \cite{Wasow} cut-off transformation around $\lambda=\lambda_0$. Note that the matrix-valued function $F_n(\lambda)$ defining the cut-off transformation should have a singularity at $\lambda=\lambda_0$, i.e. at least one of the functions $F_n(\lambda)$, $F_n^{-1}(\lambda)$ has a pole at $\lambda_0$.

Application  of this method in the discrete case is illustrated above with an example of the discrete Tzitzeica equation (see section 5).

\section{Conclusions}

The method of the formal asymptotic diagonalization of the Lax pair of nonlinear PDE proposed in \cite{Dri} is an effective tool for constructing an asymptotic representation ot the scattering problem solution and for  describing integrals of motion and higher symmetries of the associated dynamical system. The method has successfully been adopted to the discrete case in \cite{HabLOMI}, \cite{HabYang}. 

The important stage of the discrete version of the 	diagonalization method is the transformation of the Lax equation around the singular point  $\lambda=\lambda_0$ to the special form (\ref{MainSpecialForm}) where the potential is represented as a product of the regular $P_n(u,\lambda)$ and singular $Z$.

In the present article applications of the method to the discrete Tzitzeica equation and several other discrete models are studied.

\section*{Acknowledgments}

This work is partially supported by Russian Foundation for Basic Research (RFBR) grants 14-01-97008-r-Povolzhie-a  and 13-01-00070-a.

\end{document}